\documentclass[10pt,prx,twocolumn,superscriptaddress]{revtex4-2}
\usepackage[linkcolor = blue, citecolor = blue, urlcolor = blue, colorlinks = true]{hyperref}
\usepackage[usenames,dvipsnames,x11names]{xcolor}
\usepackage[mathscr]{euscript}
\usepackage[normalem]{ulem}
\usepackage{mathtools}
\usepackage{graphicx}
\usepackage{listings}
\usepackage{stmaryrd}
\usepackage{amssymb}
\usepackage{amsmath}
\usepackage{gensymb}
\usepackage{float}
\usepackage{bm}

\newcommand{\red}[1]{\textcolor{red}{#1}}

\newcommand{\traceless}[1]{\left\llbracket #1 \right\rrbracket}

\DeclareMathOperator{\ind}{ind}

\begin{document}

\title{Mechanics of axis formation in {\em Hydra}}
\author{Arthur Hernandez}
\thanks{These two authors contributed equally}
\affiliation{Instituut-Lorentz, Leiden Institute of Physics, Universiteit Leiden, P.O. Box 9506, 2300 RA Leiden, The Netherlands}
\author{Cuncheng Zhu}
\thanks{These two authors contributed equally}
\affiliation{Department of Mechanical and Aerospace Engineering, University of California San Diego, USA,}
\email{giomi@lorentz.leidenuniv.nl}
\author{Luca Giomi}
\affiliation{Instituut-Lorentz, Leiden Institute of Physics, Universiteit Leiden, P.O. Box 9506, 2300 RA Leiden, The Netherlands}
\email{giomi@lorentz.leidenuniv.nl}

\begin{abstract}
The emergence of a body axis is a fundamental step in the development of multicellular organisms. In simple systems such as {\em Hydra}, growing evidence suggests that mechanical forces generated by collective cellular activity play a central role in this process. Here, we explore a physical mechanism for axis formation based on the coupling between active stresses and tissue elasticity. We develop an active spherical shell model in which the activity of muscle fibers is described by active nematodynamics, while the elastic response of the tissue is captured by linear elasticity. We analyze the elastic deformations induced by activity-generated stresses and show that, owing to the spherical topology of the tissue, forces globally condense toward configurations in which both elastic strain and nematic defects localize at opposite poles. These mechanically selected states define either an apolar or a polar head–foot body axis. To characterize the condensed regime, we introduce a compact parameterization of the active force and flux distributions, enabling analytical predictions and direct comparison with experiments. Using this framework, we calculate experimentally relevant observables, including areal strain, lateral pressure, and normal displacements during muscular contraction, as well as the detailed structure of topological defect complexes in the head and foot regions. Together, our results identify a mechanical route by which active tissues can spontaneously break symmetry at the organismal scale, suggesting a general physical principle underlying body-axis specification during morphogenesis.
\end{abstract}

\maketitle

\section{\label{sec:introduction}Introduction}

While controlled by a myriad of biochemical pathways and regulatory networks, the organization of tissues across the various stages of embryonic development is ultimately achieved via the concerted action of a limited number of mechanical processes, whereby internally generated stresses and bending moments cooperate towards implementing the morphogenetic program. Although a complete disentanglement of biological and physical mechanisms may lie outside of the realm of possibilities, various experimental studies have recently contributed by highlighting their respective roles, signatures and limitations in model multicellular organisms. The morphogenesis of {\em Hydra}, in particular, has in recent years provided a cornucopia of biophysical behaviors where single morphogenetic pathways can be directly correlated with pivotal concepts of active living matter, including liquid crystal order and topological defects~\cite{Maroudas:2021,Fuhrmann:2024,Maroudas:2025}.

\begin{figure}[t]
\centering
\includegraphics[width=\columnwidth]{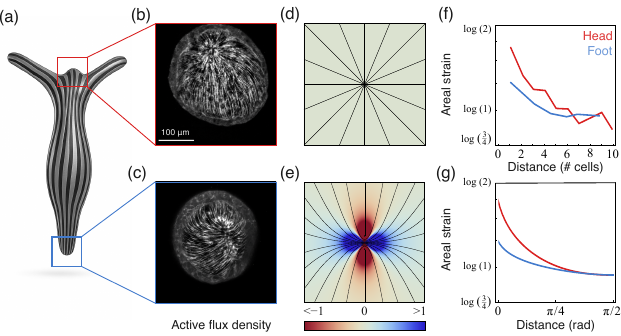}
\caption{
\label{fig:illustration}
(a) Schematic illustration of developing {\em Hydra} specimen. The black and gray stripes denote the muscle fibers of the ectoderm. (b,c) Configuration of the muscle fibers of the ectoderm 24h after excision (adapted with permission from Ref.~\cite{Maroudas:2025}). The red and blue frames highlight the future location of the head (North Pole) and foot (South Pole) respectively. Already at this stage of the regeneration process, the two configurations appear topologically distinct, with the ectodermal (endodermal) fibers at the North Pole organized around an aster (a vortex) and those at the South Pole split in two $+1/2$ disclinations. (d,e) Reconstruction of the flux density of the forces actively generated by the muscle fibers in the surrounding of a $+1$ (d) and a pair of $-1/2$ disclinations (e). In the former configuration {\em all} contractile active forces condensed at the two poles and the flux density becomes singular. (f,g) The cell areal strain -- i.e. $(A-\langle A \rangle)/\langle A \rangle$ with $A$ the cell area -- versus the distance from either the North (red) or South (blue) poles. (e) Experimental data during stretching events (adapted with permission from Ref.~\cite{Maroudas:2025}) and (f) theoretical predictions.}
\end{figure}

{\em Hydra} is a genus of small, freshwater cnidarians known for their simple tubular body structure and remarkable regenerative capabilities which allow an excised piece of tissue to regenerate into an entire functional organism. During this process, the excised tissue grows and encloses upon itself to form a spheroidal cell sheet (see Fig.~\ref{fig:illustration}). While the initial stage of this process involves a repositioning of the epithelial cells across the ectoderm and the endoderm, cell intercalation becomes negligible already after a few hours and the reorganization of the tissue concerns primarily the orientation of the muscle fibers emanating from the cells. In adult {\em Hydra}, these fibers are typically arranged in a well-ordered nematic phase, with a $+1$ disclination serving as the organizer of the muscles activating a specimen's mouth, a pair of $+1/2$ disclinations in proximity of the foot and a triplet consisting of one $+1$ and two $-1/2$ disclinations dressing each of the tentacles~\cite{Maroudas:2021} (six on average, see Ref.~\cite{Liu:1946}). Earlier in the process, on the other hand, muscle fibers are only loosely aligned and the resulting nematic texture is populated by a higher density of spurious $\pm 1/2$ disclinations and subject to periodic swelling and deswelling cycles~\cite{Beloussov:1997,SatoMaeda:1999}. Originating from the osmotic influx of water within the tissue~\cite{Benos:1977}, these oscillations drive a steady inflation of a regenerating spheroid followed by a rupture, which allows the internal fluid to be rapidly expelled~\cite{Kucken:2008}. Global coherence of nematic order is achieved in a approximatively $20$ hours, resulting in a solid tissue with fibers oriented parallel along the head-foot axis (see Fig.~\ref{fig:illustration}). 

The reorganization of muscle fibers -- of which the coarsening of $\pm 1/2$ defects is the most distinct biophysical signature -- occurs in concomitance with the localization of Wnt, i.e. a family of secreted signaling proteins that guide tissue patterning by regulating cell fate, polarity and proliferation. Specifically, the $+1$ defect marking the future location of the mouth originates from the merging of two $+1/2$ disclinations in the head region, where Wnt genes are activated in adult specimens~\cite{Broun:2005,Bode:2009}. The head-foot axis is further marked by a gradient in the lateral pressure experienced by the cells, which is maximal at the head and monotonically decreases while approaching the foot~\cite{Maroudas:2025}. How these mechanical and biochemical aspects of the morphogenesis of {\em Hydra} are causally related and how they concur to the formation of the body axis is, however, presently unknown, despite the large body of experimental data. For instance, both lateral and axial compression have been shown to alter the defect structure, possibly leading to the formation of bicephalous~\cite{Maroudas:2024} or even toroidal~\cite{Ravichandran:2025} (thus acephalous) specimens. On the other hand, Ferenc and coworkers demonstrated that Wnt signaling is sensitive to mechanical stimuli, while reducing the amount of stretching by maintaining the osmotic environment under isotonic conditions leads to a gradual decline in spheroid regenerative capacity~\cite{Ferenc:2021}. 

How this mechanism occurs in practice, however, remains elusive in spite of the theoretical effort towards deciphering the nature of the interplay between biochemical and mechanical activity. A possible pathway, proposed by Wang {\em et al.}~\cite{Wang:2023} (also see ~\cite{agam2023hydra}), relies on a morphogen {\em field} inhomogeneously distributed across the regenerating tissue. In analogy with how nematic liquid crystals reorient under an external electric or magnetic field~\cite{deGennes:1993}, spatial variations of the morphogen field could affect the local orientation of the fibers, thereby serving as a blueprint of the muscular system. This mechanism, however, requires the morphogen to be delocalized over the entire body, rather than localized at the head, and pre-patterned with a gradient in the head-foot direction. To overcome these limitations, Maroudas-Sacks {\em et al}. proposed that morphogen expression could be itself regulated based on the local strain, so that a larger strain could favor a higher concentration of the morphogen field, thus promoting a more efficient alignment of the muscular fibers in the direction of the concentration gradient~\cite{Maroudas:2025}. While the observed interplay between Wnt expression and stretching supports this hypothesis~\cite{Maroudas:2025,Ferenc:2021,Weevers:2025,zhukovsky2025hydra}, a number of questions remain instead unanswered. How can a morphogen field permeate the entire tissue while Wnt can only be detected in the head region? What is the specific regulatory pathway relating the expression of the morphogen and mechanical strain? What is the molecular origin of the proposed aligning interaction between the muscle fibers and the morphogen gradient? 

In this article, we propose an alternative mechanism facilitating axis formation in {\em Hydra}. We focus on the developmental stage observed in experiments shortly after excision, when the organism has healed itself into a spheroid but not yet established a body axis. By modeling the epithelium as an active solid, consisting
of active nematic fibers embedded in a spherical elastic shell, we show that the interplay between active nematic stresses and passive shell elasticity breaks the rotational symmetry inherited from the spherical geometry, causing the appearance of a strain gradient between opposite poles of the sphere. Remarkably, this mechanism takes place via the {\em condensation} of forces actively generated by the muscle fibers and mediated by the organization of nematic defects. Unlike previous active matter models of morphogenesis, which are either based of {\em fluid} mechanics~\cite{Vafa:2022,Al-Izzi:2023,Ho:2024,Hoffmann:2022,Hoffmann:2023,Pearce:2025}, here we specifically focus on hin-sheet elasticity and demonstrate that the necessary conditions for the occurrence of a single organizer in the head region -- in turn strictly related to the appearance of a head-foot pressure gradient -- are already available in the rich elastodynamics of Hookean shells, with activity solely providing a stress budget that passive force can redistribute.

Before entering into technical details, we provide here a concise summary of our major findings as well as an outline of this article. Our model, introduced in Sec.~\ref{sec:model}, combines three fundamental aspects of regenerating {\em Hydra}. {\em 1)} In spite of cell division and growth, cell intercalation is negligible after healing; the system is therefore a closed elastic shell able to stretch and bend under the effect of osmotic pressure and the muscle fibers. {\em 2)} These fibers, in turn, are mechanically coupled to the elastic matrix and, like in passive nematic elastomers~\cite{Lubensky:2002,Warner:2007}, can rotate when subject to a strain field. {\em 3)} Unlike in passive elastomers, however, the fibers comprising the muscular system of {\em Hydra} deliver a piece-wise constant active stress within the matrix. 

While these properties are necessary to account for the mechanical coupling between the tissue and the muscles, it is because of the spherical topology that such a coupling results in the formation of a body axis. In Sec.~\ref{sec:condensation}, we show that, for sufficiently large activity and in the presence of an aligning interaction between nematic order and strain, the active forces condense at the pole of the shell, thereby forming a body axis. This axis, in turn, can be either polar or apolar, depending on the flux of active forces emanating from the defect structures located at the poles. Specifically, in Sec.~\ref{sec:deformations} we show that the flux of the active forces across the core of the defects -- i.e. the region surrounding the defect center where the nematic order parameter drops -- can be expressed in the form
\begin{equation}\label{eq:fluxes}
\Phi_{\rm N} = 4\pi\alpha\tau\;,\qquad 
\Phi_{\rm S} = 4\pi\alpha(1-\tau)\;,	
\end{equation} 
where ${\rm N}$ and ${\rm S}$ denote the core region at the North and South pole respectively, $\alpha$ is the magnitude of the deviatoric active stress exerted by the fibers and $0\le \tau \le 1$ a number quantifying a bias in the active flux emanating from either pole. Thus, for $\tau=1/2$, $\Phi_{\rm N}=\Phi_{\rm S}=2\pi\alpha$ and the system partitions in two symmetric halves, while for $\tau > 1/2$ ($\tau<1/2$) all material fields acquire a North-South (South-North) polarity. Since the magnitude of the active forces exerted by the fibers is related to the configuration of the nematic director and this is constrained by the topology of the defect structures at each pole, a positive bias in the active flux (e.g. $\tau>1/2$) implies a core structure such as that illustrated in Fig.~\ref{fig:illustration}, comprising a $+1$ aster at the North pole (Fig.~\ref{fig:illustration}d) and a pair of antiparallel $+1/2$ disclinations at a distance $d$ from each other (Fig.~\ref{fig:illustration}e). In particular, we find that
\begin{equation}\label{eq:bias}
\tau \sim \frac{1}{2}+\left(\frac{\epsilon}{2}\right)^{4}\;,	
\end{equation}  
with $\epsilon$ the ratio between $d$ and the diameter of the core. These findings are in good qualitative agreement with experimental observations and, together with the mechanical framework detailed in the following, allow an analytical calculation of the areal strain in proximity of the defects (see Fig.~\ref{fig:illustration}f,g). Furthermore, our approach enables an explicit calculation of the surface tension of an active shell in the form
\begin{equation}\label{eq:surface_tension}
T = -P^{({\rm a})} -\frac{h^{2}}{12R^{2}}\,P^{({\rm e})}\;,
\end{equation}
where $P^{({\rm a})}$ and $P^{({\rm e})}$ are, respectively, the active and elastic contributions to the lateral pressure across the shell, $h$ the thickness and $R$ the undeformed radius.

Finally, in Sec.~\ref{sec:conclusions} we speculate about the origin of the polar bias and present a possible mechanochemical pathway leading to a stable active flux imbalance. At present, we underscore that the force-condensation mechanism underlying Eqs.~\eqref{eq:fluxes} is crucial for the emergence of this condition. Other hypothetical scenarios require a spatially extended morphogen field to be regulated by a protein confined to the head region. In contrast, our model attributes muscle-fiber alignment to a mechanical field -- specifically, the strain field -- sourced at the poles and therefore amenable to regulation by a localized control mechanism.

\section{\label{sec:model}The Model}

Despite the lack of cell intercalation, regenerating {\em Hydra} spheroids exhibit spatiotemporal fluctuations, during which the tissues comprising the endoderm and ectoderm undergo periodic inflation and collapse cycles~\cite{Beloussov:1997,SatoMaeda:1999}. In standard conditions, these oscillations have a duration of approximately $4$ hours and can therefore be treated as quasi-static with respect to the time scale associated with the rearrangement of the muscles (of the order of $10$ minutes, see e.g. Ref.~\cite{Maroudas:2025}). 

In the following, we will focus on muscular activity and imagine that, at time $t=0$, all muscle fibers in the tissue deliver a pair of equal and opposite forces of magnitude $F_{\parallel}$ along their longitudinal direction. To shed light on this process, we model the epithelium of regenerating {\em Hydra} as a spherical elastic shell of mass density $\rho$, thickness $h$ and radius $R\gg h$, whose mid-surface position is parametrized in terms of the Cartesian vector $\bm{R}=R\sin\theta\cos\phi\,\bm{e}_{x}+R\sin\theta\sin\phi\,\bm{e}_{y}+R\cos\theta\,\bm{e}_{z}$. Furthermore, let $\bm{g}_{i}=\partial_{i}\bm{R}$ be a tangent vector along the $i$-th coordinate line and $\bm{N}=\bm{R}/R$ the normal vector, so that $g_{ij}=\bm{g}_{i}\cdot\bm{g}_{j}$ is the metric tensor and $b_{ij}=-\bm{g}_{i}\cdot\partial_{j}\bm{N}=-g_{ij}/R$ the curvature tensor. A small deformation is parametrized via the mapping $\bm{R}\to\bm{R}'=\bm{R}+\delta\bm{R}$, where $\delta\bm{R}=u^{i}\bm{g}_{i}+w\bm{N}$, with $u^{i}$ and $w$ independent displacements. 

Nematic order on curved surfaces, on the other hand, can be accounted for via a covariant generalization of Landau-de Gennes theory, where $\bm{Q}=Q^{ij}\bm{g}_{i}\otimes\bm{g}_{i}$ is the order parameter tensor, with $Q^{ij}=|\Psi|(n^{i}n^{j}-g^{ij}/2)$. In the latter expression, $|\Psi|$ is the magnitude of the order parameter and $\bm{n}=n^{i}\bm{g}_{i}$ the nematic director (see e.g. Ref.~\cite{Pearce:2019} and references therein). A set of partial differential equations governing the dynamics of the fields $\delta\bm{R}$ and $\bm{Q}$, subject to the external force $\bm{f}=f^{i}\bm{g}_{i}+p\bm{N}$, can be obtained within the standard framework of shell theory and nematodynamics to give
\begin{subequations}\label{eq:eom}
\begin{gather}
\rho h\,\partial_{t}^{2}u_{i} = \nabla^{j}N_{ij}+2b_{i}^{k}\nabla^{j}M_{jk}+M_{jk}\nabla^{j}b_{i}^{k}+f_{i}\,,\\
\rho h\,\partial_{t}^{2}w  = \nabla^{i}\nabla^{j}M_{ij}-b_{ik}b_{j}^{k}M^{ij}-b_{ij}N^{ij}-p\,,\\
\partial_{t}Q_{ij}= \Gamma H_{ij}\;.
\end{gather}
\end{subequations}
Here $N_{ij}=\int {\rm d}z\,\sigma_{ij}$ is the resultant membrane stress tensor obtained upon integrating the three-dimensional {\em passive} stress $\sigma_{ij}$ across the thickness of the shell and $M_{ij}=\int {\rm d}z\,z\sigma_{ij}$ the bending moment tensor. In both definitions, $z$ denotes the distance from the mid-surface along the normal direction and spans the range $-h/2\le z \le h/2$ (see, e.g., Ref.~\cite{Niordson:1985} for an introduction). The tensor $H_{ij}$, on the other hand, is analogous the molecular tensor in nematic liquid crystals and drives the relaxation of the tensor order parameter towards a free energy minimum, with $\Gamma$ the inverse rotational viscosity~\cite{deGennes:1993}. All three tensors can be constructed upon varying the total free energy $\mathcal{F}=\int {\rm d}A\,(f^{({\rm e})}+f^{({\rm n})})$ with $f^{({\rm e})}$ and $f^{({\rm n})}$ are the elastic and nematic free energy density respectively: i.e. $N_{ij}=\delta \mathcal{F}/\delta E^{ij}$, $M_{ij}=\delta\mathcal{F}/\delta B^{ij}$ and $H_{ij}=-\traceless{\delta \mathcal{F}/\delta Q^{ij}}$. Here $E_{ij}=\delta g_{ij}/2$ and $B_{ij}=\delta b_{ij}$ are the strain and bending tensors, with $\delta g_{ij}$ and $\delta b_{ij}$ the variations in the metric and curvature tensors caused by a deformation (i.e. $g_{ij}\to g_{ij}'=g_{ij}+\delta g_{ij}$ etc.), and the operator $\traceless{\cdots}$ renders its argument traceless and symmetric. The free energy densities are instead given by
\begin{subequations}\label{eq:energy}
\begin{gather}
f^{({\rm e})} = \frac{Yh}{2(1-\nu^{2})}\Big[\nu(E^{2}+\tfrac{1}{12}h^{2}B^{2}) \notag\\
+(1-\nu)(E^{ij}E_{ij}+\tfrac{1}{12}h^{2}B^{ij}B_{ij})\Big]\;,\\[5pt]
f^{({\rm n})} =\tfrac{1}{2} L\nabla^{i}Q^{jk}\nabla_{i}Q_{jk}+\tfrac{1}{2} AQ^{ij}Q_{ij}+\tfrac{1}{4} C(Q^{ij}Q_{ij})^{2} \notag\\[5pt]
+\tfrac{1}{2} \bar{\lambda}E Q^{ij}Q_{ij}+\lambda E^{ij}Q_{ij}\;,
\end{gather}
\end{subequations}
where $E=g^{ij}E_{ij}$ and $B=g^{ij}B_{ij}$ denote covariant traces, $Y$ and $\nu$ are Young modulus and the Poisson ratio respectively, $L$ the stiffness associated with the nematic order parameter tensor and $A$ and $B$ constants setting the ``bare'' magnitude of the bulk order parameter. The parameters $\bar{\lambda}$ and $\lambda$, on the other hand, couple nematic order and strain and have been thoroughly investigated in the context of nematic elastomers, where they can be derived from a nonlinear rubber model of nematogens pinned in an elastic matrix~\cite{Lubensky:2002,Warner:2007}. Analogous phenomenological free energies have been recently used in the context of active solids~\cite{Maitra:2019,Brauns:2025}, including in earlier models of the morphogenesis of {\em Hydra}~\cite{Pearce:2020}.

Now, a central assumption of this construction is that the muscle fibers are able to deliver {\em active} forces and that the latter add to the {\em passive} forces, arising in response to elastic deformation. These forces, in turn, can be incorporated in Eqs.~\eqref{eq:eom} through the force density field $f_{i}$, representing the loads experienced by a volume element on the tangent plane of the shell, while the normal load $p$ reflects the Laplace pressure across the shell: i.e. $p=p_{\rm in}-p_{\rm out}$. For simplicity, here we consider only two active contributions resulting from the diagonal and deviatoric components of the classic active nematic stress
\begin{equation}\label{eq:active_stress}
N_{ij}^{({\rm a})}=-P^{({\rm a})}g_{ij}+\alpha Q_{ij}\;,
\end{equation}
where $P^{({\rm a})}$ is the active contribution to the pressure and $\alpha$ the magnitude of deviatoric stresses. In this case, where the forces exerted by the muscles are assumed strictly longitudinal, $P^{({\rm a})}= \rho\ell F_{\parallel}/2$ and $\alpha=-\rho\ell F_{\parallel}$, with $\ell$ the average length of the fiber~\cite{Hoffmann:2020}. Next, in Appendix \ref{sec:appendix_shell}, we show how Eqs.~\eqref{eq:eom}, \eqref{eq:energy} and \eqref{eq:active_force} can be cast in the following closed form:
\begin{subequations}\label{eq:displacementshell}
\begin{gather}
\rho h\,\partial_{t}^{2}u_{i} 
= \frac{Yh}{2(1+\nu)}\bigg[\left(\nabla^2 + \frac{1}{R^2} \right)u_{i}\notag\\
+ \frac{1+\nu}{1-\nu}\,\nabla_{i}\left(\nabla\cdot\bm{u}+\frac{2w}{R}\right)\bigg]
+ f_{i}\;,\\[10pt]
\rho h\,\partial_{t}^{2}w 
= D\left(\nabla^2+\frac{1+\nu}{R^2}\right)\left(\nabla^2+\frac{2}{R^2}\right)w\notag\\
+ \frac{Yh}{R(1-\nu)}\,\left(\nabla\cdot\bm{u}+\frac{2w}{R}\right)
+ \frac{F}{R}
- p\;,\\[10pt]
\Gamma^{-1}\partial_{t}Q_{ij} = - \left[A+\bar{\lambda}\left(\nabla\cdot\bm{u}+\frac{2w}{R}\right)+\tfrac{1}{2}B|\Psi|^{2}\right]Q_{ij} \notag\\
-\frac{\lambda}{2}\left[\nabla_{i}u_{j}+\nabla_{j}u_{i}-(\nabla\cdot\bm{u})g_{ij}\right]+L\nabla^{2}Q_{ij}\;,
\end{gather}
\end{subequations}
where $D=Yh^{3}/[12(1-\nu^{2})]$ denotes the bending rigidity of the shell. Finally, the body force $f_{i}=\nabla^{j}F_{ij}$ and the trace $F=g^{ij}F_{ij}$, in Eqs.~(\ref{eq:displacementshell}a) and (\ref{eq:displacementshell}b) respectively, are obtained from the tensor
\begin{equation}\label{eq:total_active_stress}
F_{ij}=-\left(P^{({\rm a})}-\tfrac{1}{4}\bar{\lambda}|\Psi|^{2}\right)g_{ij}+(\alpha+\lambda)Q_{ij}\;.
\end{equation}
Eqs.~\eqref{eq:displacementshell} form a closed set of partial differential equations describing how the active stresses generated by a uniform distribution of nematically ordered muscle fibers propagate through the mid-surface of a spherical shell and how the latter's elastically responds to the active and passive mechanical stimuli. We stress that, while the analysis reported here does not explicitly account for the dynamics of the Laplace pressure $p$, the latter could be systematically incorporated into the picture by modeling $p_{\rm in}$ and $p_{\rm out}$ in terms of the concentration of the medium inside and outside of regenerating spheroids (see e.g. Ref.~\cite{Kucken:2008}). 

Before proceeding with an analysis of the model, we stress that the parameters $\bar{\lambda}$ and $\lambda$, reflecting the interplay between alignment and strain, renormalize both the isotropic and deviatoric component of the active stress. For conciseness, in the remainder of this article we will incorporate these parameters directly into the definitions of $P^{({\rm a})}$ and $\alpha$, so that
\begin{subequations}\label{eq:renormalization}
\begin{gather}
\tilde{P}^{({\rm a})} = P^{({\rm a})}-\tfrac{1}{4}\bar{\lambda}|\Psi|^{2}\;,\\
\tilde{\alpha}=\alpha+\lambda\;.
\end{gather}
\end{subequations}
We stress, however, that such a renormalization can potentially affect the sign of both terms on the right-hand side Eq.~\eqref{eq:total_active_stress}, thereby altering the contractile or extensile nature of active stresses. For instance, as the muscle fibers deliver contractile stresses, we expect $P^{({\rm a})}<0$ and $\alpha>0$. The sign of $\bar{\lambda}$ and $\lambda$, on the other hand, depends on strain-alignment behavior of the system and can be either negative, if strain favors the emergence of nematic order, and positive, in the opposite case. In the following, we will drop the tilde and treat $P^{({\rm a})}$ and $\alpha$ as independent parameters, without specific constraints on their sign.

\section{\label{sec:condensation}Condensation and axis selection}

In the following, we demonstrate that the active force field undergoes a remarkable condensation phenomenon, by virtue of which active forces segregate at two opposite poles on the sphere, thereby giving rise to a body axis. The origin of this phenomenon stems from the fact that both the nematic director $\bm{n}$ and the force density $\bm{f}$, originating from its spatial variations, are constrained by the spherical topology of the shell mid-surface, thereby restricting the space of admissible solutions of Eqs.~\eqref{eq:displacementshell}. Such a static effect is further enhanced by the relaxational dynamics of the order parameter tensor, in which orientational diffusion is counteracted by the ordering field $\traceless{E_{ij}}=[\nabla_{i}u_{j}+\nabla_{j}u_{i}-(\nabla\cdot\bm{u})g_{ij}]/2$.

\begin{figure}[t]
\centering
\includegraphics[width=\columnwidth]{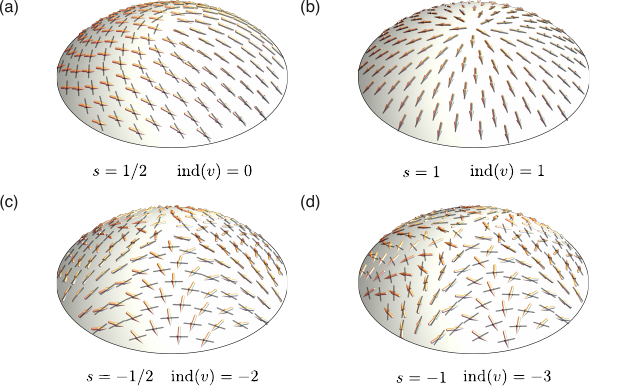}
\caption{\label{fig:vortices}Examples of force density fields (yellow arrows) originating from specific defect configurations on the sphere. The gray segments mark the local orientation of the nematic director in proximity of a nematic disclination of strength $s=\pm 1/2,\,\pm 1$.}	
\end{figure}

\subsection{\label{sec:topology}Topological constraints}

To make progress we assume that active stresses delivered by the muscle fibers are uniform in magnitude across the entire specimen, so that both $P^{({\rm a})}$ and $\alpha$ are constant. Furthermore, assuming $|\Psi|=1$ outside of the core of topological defects implies 
\begin{equation}\label{eq:active_force}
f_{i} = \alpha \nabla^{j}(n_{i}n_{j})\;.	
\end{equation}
Crucially, both sides of this relation are constrained by the spherical topology, which, in combination with Eq.~\eqref{eq:active_force}, considerably restricts the landscape of possible configurations of the force field $\bm{f}$. To illustrate this concept, let us consider a strength $s$ disclination located at the North pole of the sphere. In proximity of the defect, the nematic director takes the form: $n_{\theta}=\cos[(s-1)\phi]$ and $n_{\phi}=\sin[(s-1)\phi]$, in the orthonormal basis $\bm{e}_{\theta}=\bm{g}_{\theta}/|\bm{g}_{\theta}|$ and $\bm{e}_{\phi}=\bm{g}_{\phi}/|\bm{g}_{\phi}|$. Outside of the defect core, calculating the divergence yields $f_{\theta}=f\cos[2(s-1)\phi]$ and $f_{\phi}=f\sin[2(s-1)\phi]$, where
\begin{equation}\label{eq:force}
f = \frac{\alpha(s-1+\cos\theta)}{R\sin\theta}\;,
\end{equation}
and hence diverges like $1/r$, with $r=R\theta$ the geodesic distance from the North pole. A nematic disclination of strength $s$ introduces, therefore, a vortex $v$ of index
\begin{equation}\label{eq:indv_vs_s}
\ind(v)=2s-1\;,
\end{equation}
in the force density field generated by the muscle fibers. To compensate for this singular behavior, the magnitude $|\Psi|$ of the order parameter tensor vanishes within the core, so that $\lim_{r\to 0}\bm{f}=\bm{0}$, and the force field is everywhere smooth across the shell. Furthermore, because of the rotational symmetry of the sphere, this construction is evidently independent of the specific location of the defect. Examples of these singular configurations of both fields are shown in Fig.~\ref{fig:vortices} for different $s$ values in the range $1/2\le s \le 2$.

\begin{figure}[t]
\centering
\includegraphics[width=\columnwidth]{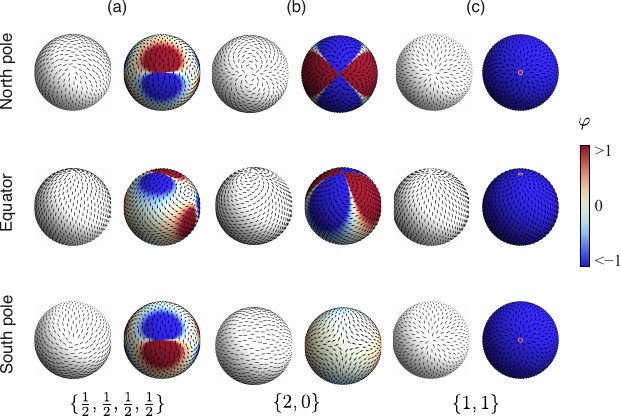}
\caption{\label{fig:textures}Examples of spherical nematic texture (left column) together with their associated active force field and flux density $\varphi$ (right column). (a) Typical configuration of spherical nematics comprising a quartet of isolated $+1/2$ disclinations. The corresponding force field consists of ten defects, of which four of index $\ind(v)=0$, colocalized with the nematic disclinations (in the center of the red/blue dumbbell-shaped features) and the remaining six, delocalized. (b) A single $+2$ disclination at the North pole. (c) Condensed state featuring two $+1$ disclinations at the poles. In this case, a uniform active flux across the entire shell is counterbalanced by an equal and opposite flux emanating from the poles.}	
\end{figure}

The existence of vortices in the active force field has crucial consequences for the organization of stress throughout the mid-surface of an active shell. Labeling $N_{\rm v}$ and $N_{\rm d}$ for the total number of vortices and defects respectively, the Poincar\'e–Hopf theorem demands
\begin{equation}\label{eq:topo_constraint}
\sum_{i=1}^{N_{\rm v}}\ind(v_{i})=\sum_{i=1}^{N_{\rm d}}s_{i}=2\;.
\end{equation}
A classic configuration satisfying this constraint consists of a quartet of $s=1/2$ disclinations (see Fig.~\ref{fig:textures}a and Refs.~\cite{Lubensky:1992,Vitelli:2006}). In this case, the force density field features ten zeros consisting of four of index $\ind(v)=0$ (co-localized with the $s=1/2$ disclinations), four of index $\ind(v)=1$ (alternating with the latter), and two of index $\ind(v)=-1$. Of the ten zeros associated with this class of configurations, therefore, four are co-localized with the nematic disclinations and six are delocalized. Analogously, during stress relaxation, a typical configuration consisting of $N_{\pm}=(N_{\rm d} \pm 4)/2$ disclinations of strength $s=\pm 1/2$ entails $N_{\rm v}=3N_{\rm d}-2$ zeros, provided $N_{\rm d}\ge 4$. Of these, $N_{+}$ of index $\ind(v)=0$ and $N_{-}$ of index $\ind(v)=-2$ are co-localized with $s=1/2$ and $s=-1/2$ disclinations respectively, and the remaining $2N_{\rm d}-2$, of which $N_{\rm d}$ of index $\ind(v)=1$ and $N_{\rm d}-2$ of index $\ind(v)=-1$, are delocalized.  

On the other hand, Eq.~\eqref{eq:topo_constraint} highlights the existence of a special class of configurations where {\em all} isolated zeros of the force density field are co-localized with nematic disclinations: i.e. $N_{\rm v}=N_{\rm d}$. By virtue of Eq.~\eqref{eq:indv_vs_s} and \eqref{eq:topo_constraint}, these configurations must feature $N_{\rm v}=N_{\rm d}=2$, thus comprise only two isolated disclinations of strength $s$ and $2-s$. Two of these configurations are shown in Fig.~\ref{fig:textures}b and \ref{fig:textures}c. In the following, we will refer to these as bipolar configurations.

\subsection{\label{sec:active_flux_appendix}Active flux condensation}

Among the bipolar configurations, that featuring two $+1$ defects located at the North and South poles of the sphere embodies a condensed state. To illustrate this concept let us introduce the {\em flux density} of the active forces: i.e. 
\begin{equation}\label{eq:flux_definition}
\varphi = \nabla^{i}\nabla^{j}F_{ij}\;.
\end{equation}
Given an arbitrary region $\mathcal{R}$ of the shell, the flux of the active forces across the boundary $\partial\mathcal{R}$ of the region is given by
\begin{equation}
\Phi_{\mathcal{R}} = \int_{\mathcal{R}} {\rm d}A\,\varphi = \oint_{\partial\mathcal{R}}{\rm d}s\,\bm{\nu}\cdot\bm{f}\;,
\end{equation}
where $\bm{\nu}$ is the outward-directed tangent-normal vector along $\partial\mathcal{R}$. Hereafter we will refer to $\Phi_{\mathcal{R}}$ as ``active flux'' and to $\varphi$ as active flux density. On a complete spherical shell, i.e. $\mathbb{S}^{2}$, global force balance requires $\Phi_{\mathbb{S}^{2}}=0$. In general, this implies that $\varphi$ varies about the neutral configuration $\varphi=0$ smoothly across the shell, see Figs.~\ref{fig:textures}a and ~\ref{fig:textures}b. This behavior is dramatically different in a bipolar configuration featuring two $+1$ defects at the North and South poles of the sphere. In this case, $\bm{n}=\bm{e}_{\theta}$ outside the cores and $\varphi=-\alpha/R^{2}$ (see Fig.~\ref{fig:textures}c and Appendix~\ref{sec:active_flux_appendix} for details). Since $\Phi_{\mathbb{S}^{2}}=\Phi_{\mathbb{S}^{2}\backslash\{{\rm N},{\rm S}\}}+\Phi_{\rm N}+\Phi_{\rm S}$, with $\mathbb{S}^{2}\backslash\{{\rm N},{\rm S}\}$ denoting the sphere punctured at both poles, this implies an equal and opposite active flux condensed within the core region: i.e.
\begin{equation}\label{eq:core_flux}
\Phi_{\rm N} + \Phi_{\rm S} = -\Phi_{\mathbb{S}^{2}\backslash\{{\rm N},{\rm S}\}} = 4\pi\alpha\left(1-\frac{a^{2}}{2R^{2}}\right)\;,	
\end{equation}
with $a$ the defect core radius and we have approximated the area of both caps as $A_{\rm cap} \approx \pi a^{2}$, under the assumption that $a\ll R$. Furthermore, in the infinitesimal core radius limit where $a/R\to 0$, Eq.~\eqref{eq:core_flux} implies that the flux density $\varphi$ attains the singular form
\begin{equation}\label{eq:delta_flux}
\varphi = \Phi_{\rm N}\,\delta(\bm{r}-\bm{r}_{\rm N})+\Phi_{\rm S}\,\delta(\bm{r}-\bm{r}_{\rm S})-\frac{\Phi_{\rm N}+\Phi_{\rm S}}{4\pi R^{2}}\;,
\end{equation}
where $\bm{r}_{\rm N}$ and $\bm{r}_{\rm S}$ indicate the position of the North and South poles respectively. Remarkably, such a condensation process selects a body axis connecting these poles. As we will detail in the next section, this body axis can be either polar or apolar depending on the magnitude of the polar fluxes $\Phi_{\rm N}$ and $\Phi_{\rm S}$.

\subsection{Head-foot symmetry breaking}

In the limit $a/R\to 0$, when the defect core radius is much smaller than the system size, Eq.~\eqref{eq:core_flux} reduces to
\begin{equation}
\Phi_{\rm N}+\Phi_{\rm S} = 4\pi\alpha\;.
\end{equation}
The simplest solution of this equation implies an equal active flux across both polar caps, so that $\Phi_{\rm N}=\Phi_{\rm S}=2\pi\alpha$. As a consequence, all material fields sourced by $\varphi$ -- including {\em all} the components of both the strain and stress tensors -- are symmetric under reflections about the equatorial plane and the body axis selected by this mechanism is, therefore, {\em apolar}. Global force balance, however, does not necessitate this additional constraint. In general, one could expect the active flux emanating by either polar region to be sensitive to endogenous and exogenous cues of both mechanical and biochemical nature -- hence $\Phi_{\rm N} \ne \Phi_{\rm S}$ -- provided global force balance is fulfilled: i.e. $\Phi_{\mathbb{S}^{2}}=0$. Crucially, once condensation has occurred, breaking the ${\rm N}\leftrightarrow{\rm S}$ symmetry only requires altering the magnitude of the fluxes sourcing elastic deformations, which, in turn, are spatially localized at the poles. 

To gain insight into this picture it is useful to reparametrize the polar flux in the form given by Eqs.~\eqref{eq:fluxes}. Thus
\begin{equation}\label{eq:alpha_tau}
\alpha = \frac{\Phi_{\rm N}+\Phi_{\rm S}}{4\pi}\;,\qquad 
\tau = \frac{\Phi_{\rm N}}{\Phi_{\rm N}+\Phi_{\rm S}}\;.
\end{equation}
Next, we show that {\em any} departure from an apolar distribution of active stresses -- hence form $\tau=1/2$ -- leaves specific ``fingerprints'' on the structure of the defects condensed at the poles. To illustrate this concept, consider two $+1/2$ disclinations located along the same meridian on the $xz-$plane at a distance $x=\pm d/2$ from the North pole, so that $\bm{n}=\cos\vartheta\,\bm{e}_{x}+\sin\vartheta\,\bm{e}_{y}$, with
\begin{equation}
\vartheta = \frac{1}{2}\,\arctan\left(\frac{y}{x-d/2}\right)+\frac{1}{2}\,\arctan\left(\frac{y}{x+d/2}\right)\;,
\end{equation}
the orientation of the nematic director in a small neighborhood of the pole. To compute the active flux emanating from the core of this defect complex, we assume that $x^{2}+y^{2}\le a^{2}$, with $a$ the radius of the core enclosing both defects, as well as the hierarchy $d \ll a \ll R$. Taking then $\epsilon=d/(2a)$ and expanding the arc-tangent yields
\begin{equation}
\vartheta = \phi + \tfrac{1}{2}\,\epsilon^{2}\sin 2\phi + \tfrac{1}{4}\,\epsilon^{4}\sin 4\phi + \tfrac{1}{6}\,\epsilon^{6}\sin 6\phi+\cdots\;.
\end{equation}
Proceeding as in Sec.~\ref{sec:topology} allows the to calculate the active force along the boundary of the core in the form 
\begin{subequations}
\begin{gather}
f_{\theta} \approx \frac{\alpha_{\rm N}}{a}\left[1+\epsilon^{2}\,\cos 2\phi - \tfrac{1}{4}\,\epsilon^{4}(1-5\cos 4\phi)\right]\;,\\
\qquad 
f_{\phi} \approx \frac{\alpha_{\rm N}}{a}\left[\epsilon^{2}\,\sin 2\phi+\epsilon^{4}\sin 4\phi\right]\;, 
\qquad
\end{gather}
\end{subequations}
where $\alpha_{\rm N}$ is the local magnitude of the deviatoric active stress at the North pole. The same construction can be repeated at the South pole, so that, integrating $\bm{f}$ along the core boundary, gives a generic expression 
\begin{equation}\label{eq:phi_p}
\Phi_{\rm P} 
= 2\pi\alpha_{\rm P}\left(1-\tfrac{1}{4}\,\epsilon^{4}\right)\;,
\end{equation}
where ${\rm P}\in\{{\rm N},{\rm S}\}$. Now, as showed in Ref.~\cite{Maroudas:2021} and reviewed in Sec.~\ref{sec:introduction}, the most commonly observed defect configuration in regenerating {\em Hydra} consists of a single $+1$ defect in the head region and a pair of bound $+1/2$ disclinations at the foot. Labeling the former as North and the latter as South pole, gives 
\begin{equation}\label{eq:phi_n_phi_s}
\Phi_{\rm N} = 2\pi\alpha_{\rm N}\;,\qquad
\Phi_{\rm S} = 2\pi\alpha_{\rm S}\left(1-\tfrac{1}{4}\,\epsilon^{4}\right)\;,
\end{equation}
from which, assuming $\alpha_{\rm N} \approx \alpha_{\rm S}$ and using Eqs.~\eqref{eq:alpha_tau}, we conclude that $\tau=4/(8-\epsilon^{4}) \approx 1/2+(\epsilon/2)^{4}$, as given by Eq.~\eqref{eq:bias}. It must be noted that in Eqs.~\eqref{eq:phi_n_phi_s}, $\epsilon$, $\alpha_{\rm N}$ and $\alpha_{\rm S}$ are not independent parameters, as the distance $d$ between $+1/2$ defects is itself determined by the magnitude of the active stress sourcing the elastic deformations. Yet, we expect that cross-correlating $\alpha$ and $\tau$ -- which, as anticipated in Fig.~\ref{fig:illustration} and detailed in Sec.~\ref{sec:deformations}, can be inferred from the configuration of the lateral pressure or the areal strain -- with a direct measurement of $\epsilon$ could provide an estimate of the individual polar active stresses $\alpha_{\rm N}$ and $\alpha_{\rm S}$. 

\subsection{\label{sec:dynamics}Condensation dynamics}

\begin{figure}
\centering
\includegraphics[width=\columnwidth]{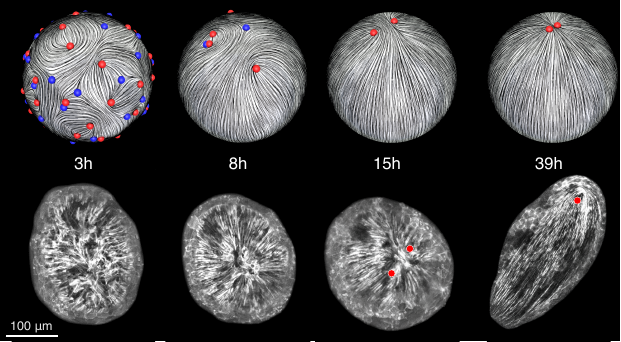}
\caption{(Top) Relaxation dynamics of an active shell obtained from a numerical integration of Eqs.~\eqref{eq:displacementshell} in the limit where $w/R=0$ and $\alpha\lambda<0$. Snapshots are shown at times $t = 0.01,\,0.1,\,0.15$ and $0.75$, from left to right. Topological defects are marked as positive half-charge (red) and negative half-charge (blue). A supplementary animation is available at \href{https://youtu.be/yoFOPmDkAcE}{this link}. (Bottom) regenerating {\em Hydra} under isotonic conditions which exhibits similar coarsening dynamics. (adapted with permission from Ref.~\cite{Maroudas:2025})}
\label{fig:nematic}
\end{figure}

In the previous subsection, we demonstrated that the active force field given by Eq.~\eqref{eq:active_force} admits a {\em condensed} state, in which like-signed active forces accumulate at the poles, while opposite-signed forces are uniformly distributed across the shell. Consistently, the nematic director marking the average direction of the muscle fibers is organized in a bipolar texture consisting of two $s=1$ disclinations at the opposite poles of the sphere. In the following, we will show that such a condensed state naturally arises from the relaxational dynamics of the nematic order parameter tensor as a consequence of the competition between orientational diffusion and strain alignment. The latter -- embodied in the term $\lambda E^{ij}Q_{ij} \sim \alpha\lambda (\bm{n}\cdot\bm{e})^{2}$ on the right-hand side of Eq.~(\ref{eq:energy}b) -- causes the nematic director to be either parallel or orthogonal to the principal strain direction $\bm{e}$. Such a direction, in turn, is either parallel or antiparallel to the active force $\bm{f}$. Thus, depending on the sign of the product $\alpha\lambda$, the nematic director is energetically favored to be parallel (for $\alpha\lambda<0$) or perpendicular (for $\alpha\lambda>0$) to the active force $\bm{f}$.

To shed light on the condensation dynamics we assume $|\Psi|=1$ outside of the defect cores and temporarily neglect out-of-plane deformations, so that the active forces exerted by the muscle fibers solely result in the build-up of in-plane strain. Furthermore, assuming inertia is negligible, in Appendix \ref{sec:active_force} we show that Eqs.~\eqref{eq:eom} can be reduced to the following 
\begin{subequations}\label{eq:active_force_eom}
\begin{gather}
\left(\nabla^2 + \frac{1}{R^2} \right)\bm{u}+ \frac{1+\nu}{1-\nu}\, \nabla (\nabla\cdot \bm{u})=-\frac{2(1+\nu)}{Yh}\bm{f}\,,\\
\Gamma^{-1}\partial_t \bm{f} = L\left(\nabla^2 +\frac{3}{R^2}\right)\bm{f} - \frac{\alpha\lambda}{2}\left(\nabla^{2}+\frac{1}{R^{2}}\right)\bm{u}\,,\!
\end{gather}
\end{subequations}
together with the global neutrality condition
\begin{equation}\label{eq:neutrality}
\int_{\mathbb{S}^{2}}{\rm d}A\,f_{i} = \int_{\mathbb{S}^{2}} {\rm d}A\,\nabla^{j}F_{ij} = 0\;.	
\end{equation}
Having neglected the nonlinear terms in Eq.~(\ref{eq:eom}c), and in particular the dynamics of the order parameter in the vicinity of defect cores, prevents us from identifying the solutions of Eqs.~\eqref{eq:active_force_eom} as genuine configurations of the displacement $\bm{u}$ and the active force field $\bm{f}$. These equations nonetheless offer a useful analogy of how strain alignment builds up during the relaxation of the nematic director. In this analogy, the longitudinally aligned configuration, where $\bm{n}=\bm{e}_{\theta}$ and active forces condense at the poles, is analogous to the trivial solution of Eqs.~\eqref{eq:active_force_eom}: i.e. $\bm{f}=\bm{0}$. In practice, this configuration can only exist in the isotropic phase, where $|\Psi|=0$ and $\bm{f}$ vanishes identically. By contrast, for finite $|\Psi|$ values, the spherical geometry constrains the nematic director into a distorted configuration, such that $\bm{f}\ne \bm{0}$. Yet, the trivial solution, for which the components $Q_{\theta\theta}=-Q_{\phi\phi}$ and $Q_{\theta\phi}=Q_{\phi\theta}$ are constant across the sphere and $\bm{f}$ is sourced solely by the intrinsic rotation of the orthonormal basis $\{\bm{e}_{\theta},\bm{e}_{\phi}\}$, provides an asymptotically exact analog of the longitudinally aligned configuration: i.e. $\lim_{R\to\infty}f\bm{e}_{\theta}=\bm{0}$, with $f$ as given in Eq.~\eqref{eq:force} when $s=1$.

To disentangle the role of the various terms in Eqs.~\eqref{eq:active_force_eom}, we first restrict the analysis to incompressible deformations, for which $\nabla\cdot\bm{u}=0$. In this case, Eqs.~\eqref{eq:active_force_eom} can be cast in the single equation
\begin{equation}\label{eq:force_dynamics}
\Gamma^{-1}\partial_{t}\bm{f} 
= L\left[\nabla^{2}+\frac{3+(1+\nu)\Lambda}{R^{2}}\right]\bm{f}\;,
\end{equation}
where $\Lambda=\alpha\lambda R^{2}/(LYh)$. Now, in the passive limit, that is when $\Lambda \to 0$, Eq.~\eqref{eq:force_dynamics} does not admit stable solutions, reflecting the lack of a uniform minimizer of the nematic free energy on the sphere. This picture is however altered by the combined effect of activity and strain alignment. Thus, for $\Lambda<\Lambda_{\rm c}=-1/(1+\nu)$, the muscle fibers align uniformly across space and $\bm{f}=\bm{0}$ (see Appendix \ref{sec:active_force} for details). Since negative $\Lambda$ values imply $\alpha\lambda<0$, condensation requires the combination of either contractile stresses and strain alignment (i.e. $\alpha>0$ and $\lambda<0$) or extensile stresses and strain anti-alignment (i.e. $\alpha<0$ and $\lambda>0$). In {\em Hydra}, where muscle fibers deliver contractile stresses, our model predicts therefore the necessity of a positive feedback between activity and strain in order for the muscles to align longitudinally. More generically, taking $\nabla\cdot\bm{u} \ne 0$ shifts the magnitude of the critical $\Lambda$ value to 
\begin{equation}\label{eq:lambda_c}
\Lambda_{\rm c}=-\frac{3+\nu}{1-\nu^{2}}\;,
\end{equation}
but without altering the qualitative picture. 

This transition bears similarities with the classic Fredericks transition in passive liquid crystals~\cite{deGennes:1993}, but, unlike the latter, the role of the ordering field -- the magnetic field in passive liquid crystals and the deviatoric strain in our model of active shells -- is here reversed. In the typical setting of the Fredericks transition, the nematic director is uniformly oriented along a fixed direction, but this configuration is unstable to rotations in the direction of the external field. By contrast, the lowest energy configuration of a spherical nematic is itself distorted, but can be rectified by the internal strain, provided the latter is sufficiently large to overweight passive forces. 

Outside of the analogy, the lack of an actual uniformly aligned configuration (i.e. $\bm{f}=\bm{0}$ everywhere) turns this continuous transition into a smooth crossover between the energy minimizing configuration of the nematic director, where the topologically required $+1/2$ disclinations are maximally spaced, to the condensed state, featuring two $+1$ defects located at the poles.

\subsection{\label{sec:numerics}Quasi-static relaxation}

In this section, we complete our analysis of the condensation process leading to the formation of an apolar body axis by means of a numerical integration of Eqs.~\eqref{eq:displacementshell} in the limit of large activity and small deformations, where $w=0$ and active forces solely result into the build up of in-plane strain. Our numerical approach is built upon the framework of complex line bundles, recently introduced by Zhu {\em et al}.~\cite{Zhu:2025} and here outlined in Appendix~\ref{sec:discretization}. 

\begin{figure}
\centering
\includegraphics[width=\columnwidth]{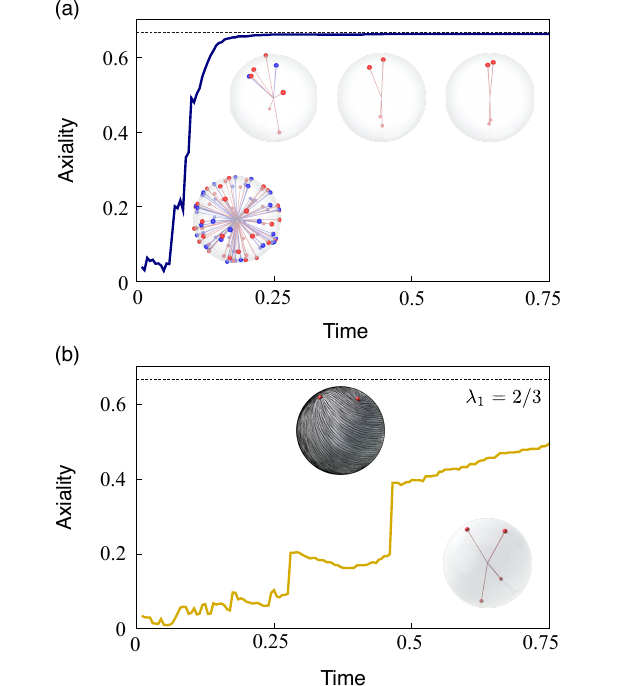}
\caption{\label{fig:axial}(a) Temporal evolution of the director's axiality for $\alpha\lambda<0$, defined as the largest eigenvalue $\lambda_1 \in [0, 2/3]$ of the polarity $\bm{P}$ in Eq.~\eqref{eq:axiality}. Snapshots of the nematic defect configurations (insets) are shown at times $t = 0.01,\,0.1,\,0.15$ and $0.75$ from left to right. Perfect axial alignment corresponds to $\lambda_1 = 2/3$ (dashed line), while the classical tetrahedral-symmetry yields $\lambda_1 = 0$. The jump in $\lambda_1$ results from discrete events of defect annihilation. (b) Same quantity for $\alpha\lambda>0$. Insets show the configurations of the director (left) and defects (right) at time $t=0.75$.}
\end{figure}

In the following, we express length in units of the sphere radius $R$ and energy in units of the orientational stiffness $L$. Furthermore, we set $|\Psi_{0}| = \sqrt{-2A/C} = 1$, $\nu = 0$ and vary the parameter $\Lambda$ in the range $-400\le \Lambda \le 400$. We initialize the nematic order parameter tensor in a highly disordered state with numerous defects (Fig.~\ref{fig:nematic}, top-left), resembling the spheroid stage in early {\em Hydra} regeneration \cite{Maroudas:2021}. To quantify the global axiality of the nematic shell, we compute the symmetric, traceless tensorial order parameter
\begin{equation}\label{eq:axiality}
\bm{P} = \frac{1}{N}\sum_{i=1}^N \left(\bm{p}_i  \otimes \bm{p}_i - \tfrac{1}{3}\,\mathbb{I}\right)\;, 
\end{equation}
where $\bm{p}_i \in \mathbb{S}^2$ denotes the position of the $i$-th defect on the sphere and $\mathbb{I}$ the three-dimensional identity tensor. Analogous to the scalar order parameter $|\Psi|$ of the nematic $\bm{Q}$-tensor (up to constant scaling), the largest eigenvalue $\lambda_1 \in [0, 2/3]$ from the spectral decomposition $\bm{P} = \sum_i \lambda_i (\bm{q}_i \otimes \bm{q}_i)$ measures the degree of axial alignment along the principal direction $\bm{q}_1$. 

\begin{figure}[b]
\centering
\includegraphics[width=\columnwidth]{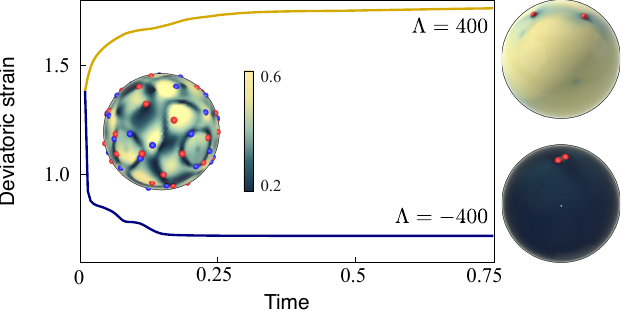}
\caption{\label{fig:strain_scalar} Relaxational dynamics of the magnitude of the mean-square deviatoric strain -- i.e. $(\traceless{E_{ij}}\traceless{E^{ij}})^{1/2}$ -- for opposite values of $\Lambda=\alpha \lambda R^{2}/(LYh)$. The inset figure shows the initial states at $t = 0$ for both cases; final states at $t = 0.75$ are shown on the right.}
\end{figure}

The typical evolution of an initially disordered configuration of the nematic director, in the presence of contractile activity and strain-alignment ($\Lambda=-400$) is shown in Fig.~\ref{fig:nematic}, together with the same process as observed by Maroudas-Sacks {\em et al}. in experiments~\cite{Maroudas:2025}. As anticipated in Sec.~\ref{sec:dynamics}, the system undergoes a coarsening regime, characterized by the rapid annihilation of $\pm 1/2$ defects, followed by a slower relaxation regime, where four topologically required $+1/2$ disclinations segregate in pairs at the poles and eventually condense in a bipolar configuration featuring two $+1$ asters. Simultaneously, the axiality parameter $\lambda_{1}$ switches from zero to nearly perfect axial alignment: i.e. $\lambda_1|_{t = 0.75} \approx 2/3$ (Fig.~\ref{fig:axial}a). This is in stark contrast with passive nematic liquid crystals, where, as detailed in Sec.~\ref{sec:topology}, the lowest energy configuration of the nematic director, subject to the global topological constraint expressed by Eq.~\eqref{eq:topo_constraint}, consists of four $s=1/2$ defects positioned as the vertices of a regular tetrahedron, for which $\lambda_{1}=0$.

By contrast, Fig.~\ref{fig:axial}b shows the evolution of axiality for the same initialization and $\Lambda = 400$. Here the relaxation towards a steady state is substantially slower than in the previous case and the asymptotic configurations features a lower degree of axiality, with $\lambda_1|_{t = 0.75} \approx 0.5$. Instead of a bipolar state, the system here relaxes towards a deformed tetrahedral configuration and active forces do not condense at the poles. The picture emerging from comparing these two scenarios is thus in qualitative agreement with the analogy presented in Sec.~\ref{sec:dynamics}, where condensation results from the linear instability of the lowest energy state. Whereas the lack of a uniform minimizer of the free energy smoothens the transition into a crossover, the critical nature of the interplay between alignment and strain leaves an imprint on the relaxation rate when $\Lambda \lessgtr 0$. 

\begin{figure}[t]
\centering
\includegraphics[width=\columnwidth]{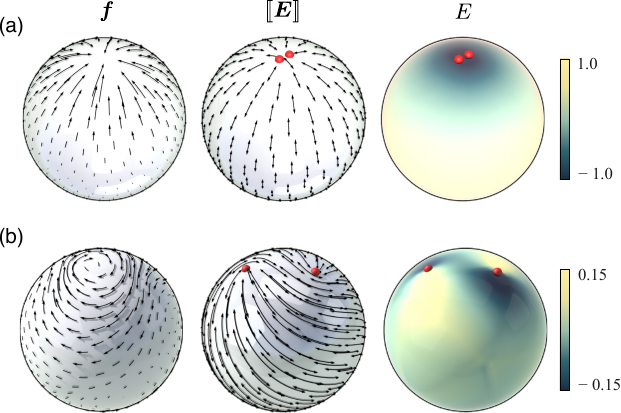}
\caption{\label{fig:stress}(a) Steady state configuration of the active force $\bm{f}$, principal component of the deviatoric strain $\traceless{\bm{E}}$ and areal strain $E$ obtained for $\Lambda=-400$. (b) Same quantities for $\Lambda=400$.}
\end{figure}

The difference between these two non-equilibrium steady states is also reflected in the distribution of strain across the mid-surface of the shell (Figs.~\ref{fig:strain_scalar} and \ref{fig:stress}). In the condensed state, deviatoric strain is minimal (Fig.~\ref{fig:strain_scalar}) and the localization of strain causes compression at poles and dilation about the equatorial region (Fig.~\ref{fig:stress}a). Moreover, the strain distribution is symmetric {\em both} axially and azimuthally. In the opposite regime, by contrast, strain is mainly deviatoric (Fig.~\ref{fig:strain_scalar}), whereas the residual areal strain is modest and less symmetric (Fig.~\ref{fig:stress}b). 

In conclusion, for either $\Lambda \lessgtr 0$, the tetrahedral symmetry characterizing the lowest energy state of spherical nematic liquid crystals is {\em explicitly} broken, as marked by the emergence of global axiality (i.e. $\lambda_{1}>0$). Consistently, the four $+1/2$ disclinations comprising the asymptotic configuration of the relaxation process localize in the polar regions, giving rise to splay-dominated (i.e. for $\Lambda<0$) or bending-dominated (i.e. for $\Lambda>0$) nematic textures. Of these two asymptotic configurations, however, only that associated to negative $\Lambda$ values exhibits a complete condensation of the active forces, concurrently to the merging of the four $+1/2$ disclination into two pairs of $+1$ defects. 

\section{\label{sec:deformations}Analysis of deformations}

Having elucidated how active forces organize on the tangent plane of our active shell model of {\em Hydra}, we conclude the investigation with an analysis of deformations. To this end, it is convenient to express the force balance condition entailed in Eqs.~(\ref{eq:eom}a) and (\ref{eq:eom}b) in terms of the elastic component of the lateral pressure $P^{(\rm e)}$ and the Gaussian curvature variation $\delta K$. These, in turn, are related to the displacement fields by
\begin{subequations}\label{eq:p_delta_k}
\begin{gather}
P^{({\rm e})} = -\frac{Yh}{2(1-\nu)}\left(\nabla\cdot\bm{u}+\frac{2w}{R}\right)\;,\\
\delta K = -\left(\nabla^{2}+\frac{2}{R^{2}}\right)\frac{w}{R}\;.
\end{gather}	
\end{subequations}
In Appendix~\ref{sec:shell} we show that a stationary configuration of these fields is obtained from
\begin{subequations}\label{eq:shell}
\begin{gather}
\!\left(\nabla^{2}+\frac{1-\nu}{R^{2}}\right)P^{({\rm e})} - \frac{1}{2}\,Yh\,\delta K= \frac{1+\nu}{2}\,\varphi\;,\\
\!DR^{2}\left(\nabla^{2}+\frac{1+\nu}{R^{2}}\right)\delta K + 2P^{({\rm e})} = -2P^{({\rm a})}-pR\;.
\end{gather}	
\end{subequations}
This form of Eqs.~(\ref{eq:eom}a) and (\ref{eq:eom}b) explicitly features the active flux $\varphi$ as a source of lateral pressure and, by virtue of their linearity, can be analytically solved once a specific configuration of the order parameter tensor $Q_{ij}$ is selected.

Before proceeding with this analysis it is useful to review two simple properties of spherical elastic shells that one can recover by taking the passive and isotropic limit of Eqs.~\eqref{eq:shell}: i.e. $F_{ij} \to 0$ and $|\Psi|\to 0$. In the absence of an externally imposed pressure difference, i.e. $p=0$, Eqs.~\eqref{eq:shell} admit the trivial solution $P^{({\rm e})}=0$ and $\delta K=0$. A non-trivial solution can instead be found when $p$ is a finite constant, in which case 
\begin{subequations}\label{eq:passive_deformations}
\begin{gather*}
P^{({\rm e})} = -\left(\frac{\gamma}{\gamma+1-\nu^{2}}\right)\frac{pR}{2}\;,\\
\delta K =-\frac{2w}{R^{3}} = -\left(\frac{1-\nu}{\gamma+1-\nu^{2}}\right)\frac{pR}{D}\;,
\end{gather*}	
\end{subequations}
where $\gamma=YR^{2}/D$ is the so-called F\"oppl-von K\'arm\'an number, expressing the relative energy cost of stretching and bending deformations~\cite{Lidmar:2003}. Thus, inflating the shell via a positive pressure difference between the exterior and the interior (i.e. $p>0$) yields an increase of the radius, thus a positive normal displacement (i.e. $w>0$) combined with a uniform extensile lateral pressure throughout the mid-surface (i.e. $P^{({\rm e})}<0$).  

\subsection{\label{sec:stretching}Stretching deformations} 

In Sec.~\ref{sec:numerics}, we showed that, in a regime where bending is negligible, the sign of $\Lambda$ determines whether the steady state configuration asymptotically approached by the relaxation process is splay- or bending-dominated, with the condensed state discussed in Sec.~\ref{sec:condensation} representing the large-activity limit of the former. As shown in Figs.~\ref{fig:strain_scalar} and \ref{fig:stress}, such a dichotomy in the organization of the nematic director is further reflected in the distribution of the residual in-plane strain, which, depending on the sign of $\Lambda$ crosses overs from primarily normal (for $\Lambda<0$) to primarily deviatoric (for $\Lambda>0$). To rationalize this property, in Appendix~\ref{sec:discretization} we show that the tensor order parameter $\bm{Q}$ admits an orthogonal decomposition into its active (force generating) and passive (divergence free) components: i.e. $\bm{Q}=\bm{Q}^{({\rm a})}+\bm{Q}^{({\rm p})}$, so that 
\[
\nabla^{j}Q_{ij}^{({\rm a})} = f_{i}\;,\qquad
\nabla^{j}Q_{ij}^{({\rm p})} = 0\;.
\]
Furthermore, $|\Psi|^{2} = |\Psi^{({\rm a})}|^{2}+|\Psi^{({\rm p})}|^{2}$, where $|\Psi^{({\rm a})}|$ and $|\Psi^{({\rm p})}|$ denote the norms of the individual components. In the absence of bending, a generic stationary solution of Eq.~(\ref{eq:eom}a) must satisfy $N_{ij} \sim -F_{ij}$, thus 
\begin{equation}
\traceless{E_{ij}} \sim -\alpha Q_{ij}^{({\rm a})}\;,
\end{equation}
from which it follows that $\mathcal{F}/L \sim - \Lambda(1+\nu)|\Psi^{({\rm a})}|^{2}$. Thus, consistently with our numerical simulations, for $\Lambda>0$ the system is energetically favored to maximize $|\Psi^{({\rm a})}|$, hence the deviatoric strain $\traceless{E_{ij}}$. By contrast, for $\Lambda<0$, minimizing $|\Psi^{({\rm a})}|$ amounts to maximize $|\Psi^{({\rm p})}|$, hence areal strain $E \sim P^{({\rm e})}$, since $|\Psi|={\rm const}$ across the shell. 

To further explore the effect of condensation on the spatial distribution of strain, we initially assume $\delta K = 0$, so that the active flux $\varphi$ is compensated exclusively by in-plane deformations. Next, parameterizing
\begin{equation}\label{eq:tau_flux}
\varphi = 4\pi\alpha\left[\tau\delta(\bm{r}-\bm{r}_{\rm N})+(1-\tau)\delta(\bm{r}-\bm{r}_{\rm S})-\frac{1}{4\pi R^{2}}\right]\;,
\end{equation}
allows finding the exact solution of Eq.~(\ref{eq:shell}a) in the condensed state, thereby offering a strategy to analytically estimate the distribution of lateral pressure and areal strain in regenerating {\em Hydra}. In the most symmetric case, where $\Phi_{\rm N}=\Phi_{\rm S}$ and $\tau=1/2$, in Appendix~\ref{sec:green_function} we show that, under these circumstances, the elastic pressure attains the form 
\begin{equation}\label{eq:symmetric_solution}
P^{({\rm e})}= \frac{\alpha(1+\nu)}{2}\sum_{\ell\in2\mathbb{N}}\frac{2\ell+1}{(1-\nu)-\ell(\ell+1)}\,P_{\ell}(\cos\theta)\;,
\end{equation}
where $2\mathbb{N}=\{2,\,4,\,6\ldots\}$ denotes the set of positive even numbers. The latter reflects the apolar distribution of the active flux sourcing in-plane deformations, as a consequence of which $P^{({\rm e})}$ is symmetric under reflections about the equatorial plane, i.e. $P^{({\rm e})}(\theta)=P^{({\rm e})}(\pi-\theta)$. For $\tau \ne 1/2$, by contrast
\begin{equation}\label{eq:non_symmetric_solution}
P^{({\rm e})}= \frac{1+\nu}{2}\sum_{\ell \in \mathbb{N}}\frac{2\ell+1}{(1-\nu)-\ell(\ell+1)}\,\alpha_{\ell}P_{\ell}(\cos\theta)\;,
\end{equation}
where we have introduced the $\ell$-dependent activity parameter
\begin{equation}\label{eq:alpha_l}
\alpha_{\ell} = \alpha\left[\tau+(-1)^{\ell}(1-\tau)\right]\;,
\end{equation}
and summation now spans the entire set of positive natural numbers, i.e. $\mathbb{N}=\{1,\,2,\,3\ldots\}$, thereby breaking the symmetry of the resulting configuration under reflections about the equatorial plane. Specifically, increasing $\tau$ above the neutral value increases the overall elastic pressure in the Southern hemisphere of the shell until, for $\tau=1$, the latter is approximatively stress-free.

\subsection{Bending deformations} 
 
If both stretching and bending are available, the same asymmetry encoded in the configuration of the elastic pressure, i.e. Eq.~\eqref{eq:non_symmetric_solution}, is inherited by the shape of the shell, whose out-of-plane deformations further reflect the uneven configuration of the active flux. Unlike stretching, however, bending deformations are additionally subject to the Laplace pressure across the shell, which, in turn, is renormalized by the active pressure $P^{({\rm a})}$.

To elucidate this coupling mechanism, as in Sec.~\ref{sec:stretching} it is instructive to start the analysis from the apolar case, where $\tau=1/2$. Assuming again axial symmetry and a uniform active and Laplace pressure (i.e. $2P^{({\rm a})}+pR={\rm const}$), one can express the solution of Eqs.~\eqref{eq:shell} in the form
\begin{subequations}\label{eq:symmetric_bending}
\begin{align}
P^{({\rm e})}
&=-\left(\frac{\gamma}{\gamma+1-\nu^{2}}\right)\frac{2P^{({\rm a})}+pR}{2} \notag \\
&+ \frac{\alpha(1+\nu)}{2}\sum_{\ell\in 2\mathbb{N}} \frac{(2\ell+1)[1+\nu-\ell(\ell+1)]}{\gamma-\nu^{2}+[1-\ell(\ell+1)]^{2}}\,P_{\ell}(\cos\theta)\;, 
\\
\delta K 
&=-\left(\frac{1-\nu}{\gamma+1-\nu^{2}}\right)\frac{2P^{({\rm a})}+pR}{D} \notag \\
&- \frac{\alpha(1+\nu)}{D}\sum_{\ell\in 2\mathbb{N}} \frac{2\ell+1}{\gamma-\nu^{2}+[1-\ell(\ell+1)]^{2}}\,P_{\ell}(\cos\theta)\;.  
\end{align}
\end{subequations}
The first constant term on the left-hand side of both equations sets the average tension and size of the active shell, whereas the second term determines the departure from a spherically symmetric distribution of both lateral pressure and curvature. Furthermore, in analogy with the Young-Laplace equation of fluid interfaces, casting Eq.~(\ref{eq:symmetric_bending}a) in the form
\begin{equation}\label{eq:young_laplace}
p = \frac{2T}{R}\;,
\end{equation}
up to $\alpha$-dependent deviatoric contributions, yields the expression of the surface tension $T$ given by Eq.~\eqref{eq:surface_tension}. In the passive limit $P^{({\rm a})}=\alpha=0$ and Eqs.~\eqref{eq:symmetric_bending} reduce to Eqs.~\eqref{eq:passive_deformations}. To best appreciate the role of deviatoric active stresses on the overall shape of the shell, one can use Eqs.~(\ref{eq:p_delta_k}b) and (\ref{eq:symmetric_bending}b) to compute the normal displacement. This gives
\begin{multline}\label{eq:apolar_w} 
\frac{w}{R^{3}}
= \left(\frac{1-\nu}{\gamma+1-\nu^{2}}\right)\frac{2P^{({\rm a})}+pR}{2D}\\
- \frac{5\alpha}{16D}\left(\frac{1+\nu}{\gamma+25-\nu^{2}}\right)(1+3\cos 2\theta)+\cdots
\end{multline}
Now, as explained in Sec.~\ref{sec:model}, all components of the active stress tensor, Eq.~\eqref{eq:active_stress}, originate from the spontaneous contraction of the muscle fibers -- thus giving $P^{({\rm a})}=-\alpha/2<0$ -- but are renormalized by the coupling between nematic order and strain: i.e. Eqs.~\eqref{eq:renormalization}. Unless the latter effect is significant enough to change the sign of $\alpha$, from Eq.~\eqref{eq:apolar_w} we conclude that a constant muscular activity changes the shape of a specimen from spherical to oblate (i.e. flatter at the poles). The latter has no consequence on the osmotic inflation cycles, since one could expect $pR \gg |P^{({\rm a})}|$, but it is likely to give rise to the short-time oscillatory dynamics consistently observed during inflation and could potentially be investigated both experimentally and theoretically. On the other hand, when subject to isotonic conditions (i.e. $p=0$), $w \sim P^{({\rm a})}$ and the spheroid is subject to a uniform shrinkage. Consistently with the observations by Ferenc {\em et al.}, such a variation in the osmotic environment has a strong impact on the organization mechanical forces and is expected to affect the performance of regeneration~\cite{Ferenc:2021}. 

\begin{figure}[t]
\centering
\includegraphics[width=\columnwidth]{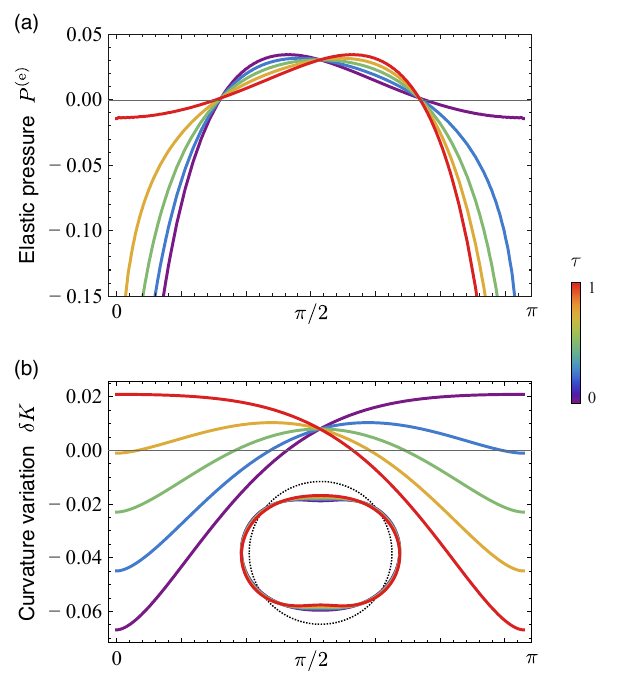}
\caption{
\label{fig:deformations}Plots of the elastic pressure $P^{({\rm e})}$ (a) and Gaussian curvature variation $\delta K$ (b) as given in Eqs.~\eqref{eq:polar_bending} for contractile activity (i.e. $\alpha>0$) and various $\tau$ values. These are indicated in panel (b). In the inset, the normal displacement $w$. The dotted circle corresponds to the circular cross-section of the undeformed spherical shell.}
\end{figure}

When $\tau>1/2$ and the symmetry under reflections about the equatorial plane is broken in favor of a polar state, assuming again a constant active and Laplace pressure throughout the shell gives
\begin{subequations}\label{eq:polar_bending}
\begin{align}
P^{({\rm e})}
&=-\left(\frac{\gamma}{\gamma+1-\nu^{2}}\right)\frac{2P^{({\rm a})}+pR}{2}+\notag \\
&+\frac{1+\nu}{2}\sum_{\ell\in \mathbb{N}} \frac{(2\ell+1)[1+\nu-\ell(\ell+1)]}{\gamma-\nu^{2}+[1-\ell(\ell+1)]^{2}}\,\alpha_{\ell}P_{\ell}(\cos\theta)\,,
\\
\delta K 
&=-\left(\frac{1-\nu}{\gamma+1-\nu^{2}}\right)\frac{2P^{({\rm a})}+pR}{D} \notag \\
&-\frac{1+\nu}{D}\sum_{\ell\in \mathbb{N}} \frac{2\ell+1}{\gamma-\nu^{2}+[1-\ell(\ell+1)]^{2}}\,\alpha_{\ell}P_{\ell}(\cos\theta)\,.
\end{align}
\end{subequations}
A plot of these functions for various $\tau$ values is reported in Fig.~\ref{fig:deformations}. The areal strain displayed in Fig.~\ref{fig:illustration} is obtained by noticing that, when subject to a lateral compression, the area $A$ of an arbitrary small portion of the shell undergoes the variation $A\to A'=A+\delta A$, with $\delta A/A = -P^{({\rm e})}/k_{B}$, with $k_{B}=Yh/[2(1-\nu)]$ the bulk modulus.

Computing the normal displacement $w$, on the other hand, requires a different strategy compared to the previous case, since the $\ell=1$ mode, which is now included in the expressions given in Eqs.~\eqref{eq:polar_bending}, is in the kernel of the operator $\nabla^{2}+2/R^{2}$ appearing in the linear variation of the Gaussian curvature, Eq.~(\ref{eq:p_delta_k}b). In Appendix~\ref{sec:displacement} we show that an expression for the $\ell=1$ mode of the normal displacement field can in this case be derived directly from Eqs.~\eqref{eq:displacementshell}, to give
\begin{multline}\label{eq:polar_w} 
\frac{w}{R^{3}}
= \left(\frac{1-\nu}{\gamma+1-\nu^{2}}\right)\frac{2P^{({\rm a})}+pR}{2D}\\
+ \frac{2R^{2}(1+\nu)}{Yh}\,f_{1}^{(1)}\cos\theta+\cdots
\end{multline}
where $f_{1}^{(1)}\sim -\alpha_{1}$ is the amplitude of $\ell=1$ mode in the vector spherical harmonics expansion of the force density field $\bm{f}$. Polarity, therefore, increases the flatness around the North pole, where the active flux is stronger (see inset of Fig.~\ref{fig:deformations}b).

\section{\label{sec:conclusions}Discussion and Conclusions}

We have proposed an active shell framework to explain the body axis selection in {\em Hydra} as an example of global spontaneous symmetry breaking. This framework is grounded in three physical properties of the epithelium of {\em Hydra}: i.e. structural integrity, nematic order across the actin cytoskeleton and contractile activity. We demonstrated that, when combined with the spherical topology, which is characteristic of early embryonic development, these properties engender a striking condensation phenomenon, in which like-signed active forces accumulate at the poles, while opposite-signed forces are uniformly distributed across the shell. This process identifies a body axis, connecting the two opposite poles, along which muscle fibers align. Such an axis, in turn, can be either polar or apolar, depending on whether the fluxes of the active forces emanating from the poles -- $\Phi_{\rm N}$ and $\Phi_{\rm S}$ -- are equal or different. Unlike in other mechanochemical models of {\em Hydra}~\cite{Wang:2023,Maroudas:2025,Weevers:2025}, this mechanism does not require the existence of a spatially extended morphogen field, but assigns an analogous regulatory role to strain, whose configuration over the scale of the entire organism is determined by the organization of the active forces at the poles.

Our approach introduces an especially convenient strategy for parameterizing the distribution of active forces and their fluxes in the condensed state, thus opening the door to the analytical calculation of multiple quantities of general interest. These include the spatial configuration of the areal strain, lateral pressure and normal displacements resulting from muscular contraction, as well as the fine-scale structure of the complexes of topological defects in the head and foot region. In combination with a direct experimental characterization of morphology and nematic order across specimens, our approach could pave the way to the estimation of various material parameters -- such as the mean active stress $\alpha=(\Phi_{\rm N}+\Phi_{\rm S})/(4\pi)$ and the polar bias $\tau=\Phi_{\rm N}/(\Phi_{\rm N}+\Phi_{\rm S})$ -- and shed light on the interplay between mechanical response and biochemical regulation in {\em Hydra}. Furthermore, while the present analysis focuses on the quasi-static properties, the framework is more general and can be used to investigate time-dependent aspects of the regeneration process, such as inertial effects and the persistence of structural features in between inflation/deflation cycles. 

\begin{figure}[b]
\centering
\includegraphics[width=\columnwidth]{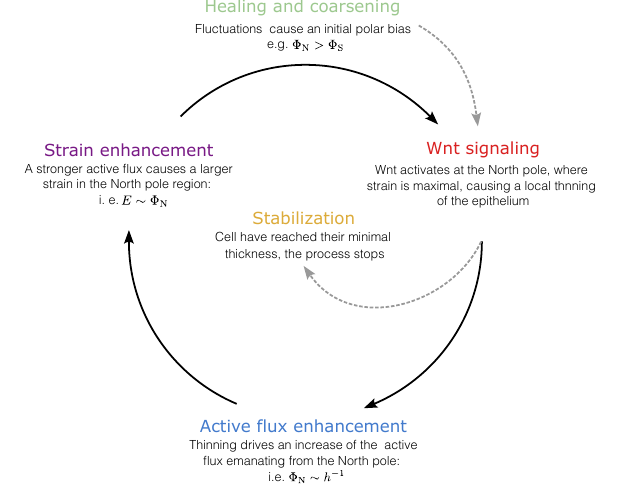}
\caption{\label{fig:feedback}A possible feedback loop governing the reinforcement of the initial polar bias. As demonstrated in Ref.~\cite{Weevers:2025}, Wnt activates where the strain is maximal, leading to a reduction of the local thickness of the epithelium. The latter enhances the two-dimensional active flux emanating from the North pole, thereby enhancing the local strain, hence the polar bias. The process stops once the cells have reached thir minimal thickness, thus resulting in a mechanochemically stable North-South axis.}
\end{figure}

While a thorough investigation of the origin of the polar bias requires a separate study, some considerations are in order. Since the polar fluxes $\Phi_{\rm N}$ and $\Phi_{\rm S}$ sourcing the distribution of strain originate themselves from the concerted action of multiple muscle fibers -- i.e. $\Phi_{\rm P}=\int_{\rm P} {\rm d}A\,\varphi$, with ${\rm P}\in\{{\rm N},{\rm S}\}$ denoting the core of a polar defect -- the probability of a perfectly symmetric balance of active flux at the poles is evidently slim. By contrast, we expect a polar bias to be present already after the initial coarsening phase of self-healing, when the excised tissue has closed itself into a spheroid and most of the topological defects have annihilated. Such an initial bias, on the other hand, could be reinforced by Wnt signaling, thus evolving into a mechanochemically stable head-foot axis. Important steps towards elucidating this regulatory pathway have been recently reported by Weevers {\em et al}. in Ref.~\cite{Weevers:2025}. Combining mechanical manipulations with a time series RNA sequencing of regenerating spheroids, Weevers {\em et al}. unveiled three fundamental aspects of the mechanochemical cross-talk in {\em Hydra}: {\em 1)} Wnt3 ligands cause a decrease in surface tension, as revealed by micropipette aspiration experiments; {\em 2)} Wnt3 activation, in turn, is sensitive to mechanical stimuli and more prominent where the tissue is subject to higher stretching; {\em 3)} in these regions, the thickness of the epithelium is substantially smaller than across the entire spheroid, as indicated by a direct imaging of the Wnt-associated protein $\beta$-catenin. To rationalize these observations, Weevers {\em et al}. postulated the existence of a positive feedback loop correlating the Young modulus of the tissue and the level of expression of Wnt, with the latter serving as a blueprint for other anatomical features. 

Our analysis, however, suggests an alternative scenario, which does not require a spatially extended morphogen field and where the role of Wnt regulation is comparable to that of muscular activity. First, whether or not Wnt ligands affect the Young modulus of the tissue, from Eq.~\eqref{eq:surface_tension} and Fig.~\ref{fig:deformations} we see that, near the poles, the magnitude of surface tension is dominated by the elastic pressure, so that $T \sim h^{2}$. The reduction of surface tension reported by Weevers {\em et al}. in samples of Wnt3-overexpressing spheroids, could then be directly ascribed to the aforementioned decrease of the cell thickness $h$. The latter, on the other hand, is expected to effectively increase the two-dimensional active flux $\Phi_{\rm P}$ emanating from the Wnt-rich pole of the mid-surface. In general, $\Phi_{3{\rm D}}=h\Phi_{\rm P}$, where $\Phi_{3{\rm D}}$ is the flux across the pillbox surface obtained upon extruding the two-dimensional core through the thickness of the shell, from the endoderm to the ectoderm. Thus, assuming the total force exerted by the muscle fibers independent of the thickness of the epithelium, implies $\Phi_{\rm 3D}={\rm const}$ and $\Phi_{\rm P}\sim h^{-1}$. A local thinning of the tissue, therefore, results in a stronger {\em effective} two-dimensional active flux, thus a larger strain across the North pole region and a higher alignment of the muscle fibers, as reflected by the fine-scale structure of the defects. As Wnt activates in response to strain, the latter process entails a positive feedback loop that reinforces the initial polar bias towards establishing a stable head-foot axis. The process eventually stops or becomes irrelevant once the cells have reached the minimal thickness allowed by the osmotic environment. 

Fig.~\ref{fig:feedback}) summarizes the proposed mechanochemical regulatory pathway. The latter could be experimentally tested by reconstructing both active fluxes $\Phi_{\rm N}$ and $\Phi_{\rm S}$ via optical imaging of the muscle fibers while locally manipulating the thickness of the tissue.

\acknowledgments

We acknowledge useful discussions with Kinneret Keren, Marko Popovic, Fridtjof Brauns, Cristina Marchetti and Zhibang Yao during the preparation of this article.

\appendix

\section{\label{sec:appendix_shell}Derivation of Eqs.~\eqref{eq:displacementshell}}

A thorough introduction to elastic shells was given by Niordson in Ref.~\cite{Niordson:1985}. For sake of completeness, in the following, we provide a concise derivation of tangential and normal force balance given by Eqs.~(\ref{eq:displacementshell}a) and~(\ref{eq:displacementshell}b), respectively. Using the classic parametrization of the sphere given in Sec.~\ref{sec:model} allows expressing the components of the metric tensor in the form: $g_{\theta\theta} = R^{2}$, $g_{\phi\phi} = R^{2}\sin^{2}\theta$, and $g_{\theta\phi} = g_{\phi\theta} = 0$, while the curvature tensor is given by $b_{ij}=-g_{ij}/R$. Using this in Eqs.~\eqref{eq:eom} one readily finds
\begin{subequations}\label{eq:eom_appendix}
\begin{gather}
\rho h\,\partial_{t}^{2}u_{i} = \nabla^{j}\left(N_{ij}-\frac{2}{R}\,M_{ij}\right) + f_{i}\;,\\
\rho w\,\partial_{t}^{2}w = \nabla^{i}\nabla^{j}M_{ij}-\frac{M}{R^{2}}+\frac{N}{R}-p\;.
\end{gather}
\end{subequations}
The stretching and bending tensors, on the other hand, are given at the linear order in the displacements by
\begin{gather*}
E_{ij} = \tfrac{1}{2}(\nabla_{i}u_{j}+\nabla_{j}u_{i})+\frac{w}{R}\,g_{ij}\;,\\
B_{ij} = \nabla_{i}\nabla_{j}w-\frac{1}{R}\,(\nabla_{i}u_{j}+\nabla_{j}u_{i})-\frac{w}{R^{2}}\,g_{ij}\;.
\end{gather*}
For sake of convenience, one can incorporate in the vector field $f_{i}$ all the forces that are not directly sourced by $E_{ij}$ and $B_{ij}$, so that $f_{i}=\nabla^{j}F_{ij}$, where the tensor $F_{ij}$, whose explicit expression is given in Eq.~\eqref{eq:total_active_stress}, accounts for stresses originating from both activity and strain alignment. The stress and bending moment tensors, in turn, are related to the strain tensor $E_{ij}$ and the bending tensor $B_{ij}$ by the Hookean constitutive equations
\begin{subequations}\label{eq:hooke}
\begin{gather}
N_{ij} = \frac{Yh}{1-\nu^{2}}\left[(1-\nu)E_{ij}+\nu E g_{ij}\right]\;,\\
\quad M_{ij} = D\left[(1-\nu)B_{ij}+\nu B g_{ij}\right]\;,\quad
\end{gather}
\end{subequations}
Now, Eqs.~\eqref{eq:eom_appendix} can be substantially simplified by taking advantage of the possibility of taking alternative measures of both stretching and bending~\cite{Niordson:1985}. This amounts to replacing $N_{ij}$ and $B_{ij}$ with
\begin{gather*}
\tilde{N}_{ij} = N_{ij}-\frac{2}{R}\,M_{ij}\;,\\
\tilde{B}_{ij} = B_{ij}+\frac{2}{R}\,E_{ij}=\nabla_{i}\nabla_{j}w+\frac{w}{R}\,g_{ij}\;.
\end{gather*}
It is simple to verify that these augmented stress and bending tensors satisfy the condition
\[
N^{ij}\delta E_{ij} + M^{ij} \delta B_{ij} = \tilde{N}_{ij}\delta E_{ij} + {M}_{ij}\delta \tilde{B}_{ij}\;,
\]
and are therefore equivalent to $N_{ij}$ and $B_{ij}$ by virtue of the principle of virtual work. 

These manipulations allow us to cast Eqs.~\eqref{eq:eom_appendix} in the simpler form
\begin{subequations}\label{eq:eom_appendix_transformed}
\begin{gather}
\rho h\,\partial_{t}^{2}u_{i} = \nabla^{j}(\tilde{N}_{ij}+F_{ij})\;,\\
\rho h\,\partial_{t}^{2}w = \nabla^{i}\nabla^{j}M_{ij}+\frac{M}{R^{2}}+\frac{1}{R}\,(\tilde{N}+F) - p\;.
\end{gather}
\end{subequations}
where the stress and bending moment tensors are given by
\begin{gather*}
\tilde{N}_{ij} = 
\frac{Yh}{1-\nu^{2}}\,\bigg[
\frac{1-\nu}{2}\,\left(\nabla_{i}u_{j}+\nabla_{j}u_{i}+\frac{2w}{R}\,g_{ij}\right)\notag \\
+\nu g_{ij} \left(\nabla\cdot\bm{u}+\frac{2w}{R}\right)\bigg]\;,\\
M_{ij} = D\bigg[(1-\nu)\left(\nabla^{i}\nabla^{j}w+\frac{w}{R^{2}}\,g_{ij}\right) \notag \\
+\nu g_{ij}\left(\nabla^{2}+\frac{2}{R^{2}}\right)w\bigg]\;.
\end{gather*}
From these expressions, the traces $\tilde{N}$ and $M$ can be straightforwardly calculated in the from
\begin{gather*}
\tilde{N} = \frac{Yh}{1-\nu}\left(\nabla\cdot\bm{u}+\frac{2w}{R}\right)\;,\\
M = D(1+\nu)\left(\nabla^{2}+\frac{2}{R^{2}}\right)w\;.
\end{gather*}
To compute $\nabla^{j}\tilde{N}_{ij}$ and $\nabla^{i}\nabla^{j}M_{ij}$ we first recall the commutator of the covariant derivatives of a generic rank$-p$ tensor field $T_{k_{1}k_{2}\cdots\,k_{p}}$: i.e.
\begin{multline}\label{eq:commutator}
[\nabla_{i},\nabla_{j}]T_{k_{1}k_{2}\cdots\,k_{p}}
= R_{k_{1}ji}^{l}T_{lk_{2}\cdots\,k_{p}}\\
+ R_{k_{2}ji}^{l}T_{k_{1}l\cdots\,k_{p}}
\cdots
+ R_{k_{p}ji}^{l}T_{k_{1}k_{2}\cdots\,l}\;,
\end{multline}
where $R_{ijk}^{l}$ is the Riemann tensor. For surface of Gaussian curvature $K$, the latter is given by
\begin{equation}\label{eq:riemann}
R_{ijk}^{l} = K(\delta^{l}_{j}g_{ik}-g_{ij}\delta^{l}_{k})\;.
\end{equation}
Eqs.~\eqref{eq:commutator} and \eqref{eq:riemann}, in turn, can be used to demonstrate the following commutation relations for covariant vectors $\bm{v}=v^{i}\bm{g}_{i}$: i.e. 
\begin{equation}\label{eq:div_commutator}
[\nabla^{i},\nabla_{j}]v_{i} = Kv_{j}\;.
\end{equation}
On the sphere, where $K=1/R^{2}$, this yields the following expressions
\begin{gather*}
\nabla^{j}\nabla_{i}u_{j} = \nabla_{i}\nabla\cdot\bm{u}+ \frac{u_{i}}{R^{2}}\;,\\
\nabla^{i}\nabla^{j}\nabla_{i}\nabla_{j}w = \nabla^{2}\left(\nabla^{2}+\frac{1}{R^{2}}\right)w\;,
\end{gather*}
Substituting in Eqs.~\eqref{eq:eom_appendix_transformed} finally gives Eqs.~\eqref{eq:eom}. 

\section{\label{sec:active_flux_appendix}Active flux in bipolar configurations}

The generic expression of the force density field associated with the fully localized bipolar configurations shown in Fig.~\ref{fig:textures} is given in Sec.~\ref{sec:condensation} in the form
\begin{equation}\label{eq:generic_force}
\bm{f} = f_{+}\cos[2(s-1)\phi]\,\bm{e}_{\theta}+f_{-}\sin[2(s-1)\phi]\,\bm{e}_{\phi}\;,
\end{equation}
where the coefficients $f_{\pm}$ are given by
\begin{equation}
f_{\pm} = \frac{\alpha|\Psi|(s-1+\cos\theta)}{R\sin\theta} \pm \frac{1}{R}\,\partial_{\theta}|\Psi|\;.
\end{equation}
Outside of the core of the defects, where $|\Psi|=1$, $f_{\pm}\to f$, thereby recovering the expression given in Eq.~\eqref{eq:force}. For $s=1$, Eq.~\eqref{eq:generic_force} reduces to
\begin{equation}
\bm{f} = \frac{\alpha\cot\theta}{R}\,\bm{e}_{\theta}\;, 
\end{equation}
from which, computing the divergence, gives
\begin{equation}
\nabla\cdot\bm{f}
= \frac{1}{R\sin\theta}\,\partial_{\theta}(f_{\theta}\sin\theta)+\frac{1}{R\sin\theta}\,\partial_{\phi}f_{\phi}=-\frac{\alpha}{R^{2}}\;. 
\end{equation}
Thus, integrating the divergence over the entire sphere and using $\Phi_{\mathbb{S}^{2}\backslash\{{\rm N},{\rm S}\}}+\Phi_{{\rm N}}+\Phi_{{\rm S}}=0$, gives Eq.~\eqref{eq:core_flux}, hence
\begin{equation}\label{eq:core_flux_appendix}
\Phi_{{\rm N}}+\Phi_{{\rm S}} = 4\pi\alpha+\mathcal{O}[(a/R)^{2}]\;. 
\end{equation}
The same result can be obtained calculating the flux of $\bm{f}$ across the cores. Taking $a=R\theta_{\rm c}$, with $\theta_{\rm c}$ the angular latitude of the boundary of the northern cap, and taking $\bm{\nu}=\bm{e}_{\theta}$, gives 
\begin{equation}
\Phi_{{\rm N}} = \oint_{\partial{\rm N}} {\rm d}s\,\bm{\nu}\cdot\bm{f} 
= 2\pi\alpha\,\frac{a}{R \tan \theta_{\rm c}} 
\xrightarrow{\theta_{\rm c}\to 0} 2\alpha\pi\;.
\end{equation}
The same result holds on the South pole, from which one finds Eq.~\eqref{eq:core_flux_appendix}. In the general case of a disclination of strength $s$ located at the North pole, one has
\begin{equation}
\oint_{\partial{\rm N}} {\rm d}s\,\bm{\nu}\cdot\bm{f} 
= a f(\theta_{\rm c} )\int_{0}^{2\pi} {\rm d}\phi\,\cos[2(s-1)\phi]=0\;, 	
\end{equation}
for any $s \ne 1$. The bipolar configuration featuring two $s=1$ asters at the poles is, therefore, the only configuration of nematic defects on the sphere resulting from the condensation of active forces.

\section{\label{sec:active_force}Active force dynamics}

To derive Eqs.~\eqref{eq:active_force_eom}, we approximate $w/R = 0$ and assume $|\Psi|=1$ outside of the defect core. Multiplying both sides of Eq.~(\ref{eq:eom}c) by $\alpha$ and taking the divergence gives then
\begin{equation}\label{eq:dqdt_appendix}
\Gamma^{-1} \partial_{t} f_{i} = L\nabla^{j}\nabla^{2}Q_{ij}-\alpha\lambda\nabla^{j}\traceless{E_{ij}}\;,
\end{equation}
where $f_{i}=\alpha\nabla^{j}Q_{ij}$ and $\traceless{E_{ij}}$ denotes the traceless part of the strain tensor: i.e. 
\[
\traceless{E_{ij}} = \frac{1}{2}\left[\left(\nabla_{i}u_{j}+\nabla_{j}u_{i}\right)-\left(\nabla\cdot\bm{u}\right)g_{ij}\right]\,.
\]
Eq.~(\ref{eq:active_force_eom}a) follows straightforwardly from Eq.~(\ref{eq:eom}a), while Eq.~(\ref{eq:active_force_eom}b) is obtained by taking into account that, by virtue of Eqs.~\eqref{eq:commutator} and \eqref{eq:riemann}
\[
[\nabla^{j},\nabla^{2}]Q_{ij} = K\nabla^{j}Q_{ij}+2\nabla^{j}(KQ_{ij})\;.
\]
From this and Eq.~\eqref{eq:commutator}, taking $K=1/R^{2}$ gives
\begin{subequations}\label{eq:active_force_commutators}
\begin{gather}
\nabla^{j}\nabla^{2}Q_{ij} = \left(\nabla^{2}+\frac{3}{R^{2}}\right)\nabla^{j}Q_{ij}\;,\\
\nabla^{j}\traceless{E_{ij}} = \left(\nabla^{2}+\frac{1}{R^{2}}\right)\frac{u_{i}}{2}\;.
\end{gather}
\end{subequations}
Finally, replacing Eqs.~\eqref{eq:active_force_commutators} in Eq.~\eqref{eq:dqdt_appendix} readily yields Eqs.~(\ref{eq:active_force_eom}b).

Next, to assess the stability of the trivial solution of Eq.~\eqref{eq:force_dynamics}, we decompose $\bm{f}$ in terms of vector spherical harmonics -- i.e. $\bm{\Psi}_{\ell}^{m} = R\nabla Y_{\ell}^{m}$ and $\bm{\Phi}_{\ell}^{m} = \bm{R}\times \nabla Y_{\ell}^{m}$, where $Y_{\ell}^{m}=Y_{\ell}^{m}(\theta,\phi)$ is a spherical harmonic of order $(\ell,m)$. Thus 
\begin{equation}\label{eq:force_Psi_Phi}
\bm{f} = \sum_{\ell=1}^{\infty}\sum_{m=-\ell}^{\ell}\left(f_{\ell m}^{(1)}\bm{\Psi}_{\ell}^{m}+f_{\ell m}^{(2)}\bm{\Phi}_{\ell}^{m}\right)\;.
\end{equation}
The basis vectors $\bm{\Psi}_{\ell}^{m}$ and $\bm{\Phi}_{\ell}^{m}$ are mutually orthogonal and characterized by the following differential properties 
\begin{align*}
\nabla\cdot\bm{\Psi}_{\ell}^{m} &= -\frac{\ell(\ell+1)}{R}\,Y_{\ell}^{m}\,, 
&&\nabla^{2}\bm{\Psi}_{\ell}^{m}   = -\frac{\ell(\ell+1)}{R^{2}}\,\bm{\Psi}_{\ell}^{m}\,,\\
\nabla\cdot\bm{\Phi}_{\ell}^{m} &= 0\,,
&&\nabla^{2}\bm{\Phi}_{\ell}^{m}   = -\frac{\ell(\ell+1)}{R^{2}}\,\bm{\Phi}_{\ell}^{m}\,.
\end{align*}
From these and Eq.~\eqref{eq:force_Psi_Phi} and Eq.~\eqref{eq:force_dynamics} it is possible to calculate the growth rate of the individual modes. That is
\[
\partial_{t}f_{\ell m}^{(n)} = D_{\ell}^{(n)}f_{\ell m}^{(n)}\;,
\]
with $n=1,\,2$ and 
\[
D_{\ell}^{(n)} = \frac{D_{\rm r}}{R^{2}}\left[3+(1+\nu)\Lambda-\ell(\ell+1)\right]\;,
\]
with $D_{\rm r} = \Gamma L$ the rotational diffusion coefficient. Evidently, $\bm{f}=\bm{0}$ is a stable solution of Eq.~\eqref{eq:force_dynamics} only when $D_{\ell}^{(n)}<0$. Since $\ell=1$ is the first mode to become unstable, taking $D_{1}=0$ readily yields $\Lambda_{\rm c}=-1/(1+\nu)$. Similarly, lifting the constraint of incompressibility and expanding 
\begin{equation}\label{eq:u_lm}
\bm{u} = \sum_{\ell=1}^{\infty} \left(u_{\ell m}^{(1)}\bm{\Psi}_{\ell}^{m}+u_{\ell m}^{(2)}\bm{\Phi}_{\ell}^{m}\right)\;,
\end{equation}
gives, after simple algebraic calculations 
\begin{gather*}
u_{\ell m}^{(n)} = C_{\ell}^{(n)} f_{\ell m}^{(n)}\;,\\
\partial_{t}f_{\ell m}^{(n)} = D_{\ell}^{(n)} f_{\ell m}^{(n)}\;.
\end{gather*}
The coefficients $C_{\ell}^{(n)}$ and $D_{\ell}^{(n)}$ are given by
\begin{gather*}
C_{\ell}^{(n)} = \frac{2R^{2}/(Yh)(1-\nu^{2})}{\ell(\ell+1)[2+(1-n)(1+\nu)]-(1-\nu)}\;,\\[10pt]
D_{\ell}^{(n)} = \frac{D_{\rm r}[3-\ell(\ell+1)]}{R^{2}}-\frac{\alpha\lambda\Gamma}{2R^{2}}\,[1-\ell(\ell+1)]\,C_{\ell}^{(n)}\;.
\end{gather*}
For $\ell=1$, in particular, this gives
\begin{subequations}
\begin{gather}
D_{1}^{(1)} = 	\frac{D_{\rm r}}{R^{2}}\left[1+\left(\frac{1-\nu^{2}}{3+\nu}\right)\Lambda\right]\;,\\[10pt]
D_{1}^{(2)} = 	\frac{D_{\rm r}}{R^{2}}\left[1+(1+\nu)\Lambda\right]\;.
\end{gather}
\end{subequations}
Thus, for $-1 < \nu < 1$ -- thereby including both standard and weakly auxetic behavior –- the $\bm{\Psi}_{1}^{m}$ mode becomes unstable for $\Lambda<\Lambda_{\rm c}$, with $\Lambda_{\rm c}$ given in Eq.~\eqref{eq:lambda_c}. 

\section{\label{sec:discretization}Complex line bundle discretization scheme}

In this Appendix we outline the complex bundle formulation of Eqs.~\eqref{eq:displacementshell}. This approach forms the foundation of the numerical discretization framework developed in Ref.~\cite{Zhu:2025}, which applies to surfaces of arbitrary geometry and topology. In addition, the use of bundle formalism helps clarify the orthogonal decomposition in Sec.~\ref{sec:stretching}, as well as help draw connections to the problems in the mathematical elasticity literature regarding the Saint-Venant compatibility problem~\cite{Ciarlet:2005,Marsden:1994} and existence of stress potentials~\cite{Kupferman:2022,Kupferman:2025}.

\subsection{Shell model via complex line bundles}

On a surface $M$, let $\bm 1 \in \Gamma_{U}(TM)$ with $|\bm 1| = 1$ be a local unit basis vector field defined over a region $U \subset M$. Its corresponding nematic basis is defined as the equivalence class $[\bm 1] \in \Gamma_U(TM / \mathbb{Z}_2)$, identifying $\mathbb{I} \sim -\bm 1$. Using the basis, vector and nematic fields can be expressed as complex-valued functions. A displacement field 
\begin{align}
\textstyle
\bm u = \hat u \bm 1 \in \Gamma_U(TM)\;,
\end{align}
corresponds to $\hat u \in \mathbb C$. Likewise, a nematic field $\Psi = \hat \Psi [\bm 1]$ is represented by the complex order parameter 
\begin{align}
\textstyle
\hat \Psi = |\Psi| e^{\mathrm i 2 \theta} \in \mathbb C\;.
\end{align}
This complex representation is isomorphic to the symmetric, traceless tensor form $\bm Q = |\Psi| (\bm n \otimes \bm n - \bm g / 2) \in \Gamma(TM \otimes TM)$, where the director is $\bm n =  e^{\mathrm i \theta} \bm 1$. The bijective map between symmetric traceless 2-tensor and nematic director --- known as the \emph{Veronese map} --- is denoted 
\begin{align}
\textstyle
\mathcal V: \Gamma(TM/\mathbb Z_2) \rightarrow \Gamma(TM \otimes TM), \quad \mathcal{V}: \Psi \mapsto \bm Q\;.
\end{align}
We introduce the covariant derivatives on tangent and nematic bundles within the complex framework. The gradient of a scalar function $f \in \Gamma(M; \mathbb R)$ on $M$ is given by $\operatorname{grad} f \coloneqq ({\rm d}f)^\sharp \in \Gamma(TM)$. For the tangent bundle, the covariant derivative of $\bm u$ is given by 
\begin{align}
\nabla \bm u = ({\rm d}\hat u + \mathrm i \omega  \hat u) \bm 1\;,
\end{align}
where the connection 1-form $w$ is defined by $\nabla \bm 1 \eqqcolon \mathrm i w \bm 1$. For the nematic bundle, we denote the covariant derivative by a subscript and write 
\begin{align}
\nabla_2 \Psi = ({\rm d} \hat \Psi + \mathrm i 2 \omega  \hat \Psi) [\bm 1]\;. 
\end{align}
The nematic connection $\nabla_2 [\bm 1] = \mathrm i 2\omega [\bm 1]$ accounts for the doubled rotational speed associated with nematic symmetry. The curvature of the connection measures the holonomy of parallel transport and is given by 
\begin{align}
{\rm d}A\,K  = -{\rm d}\omega\;.
\end{align}
This complex-valued formulation parallels the classical Ginzburg-Landau theory, where $\omega$ plays the role of the magnetic vector potential and $K$ corresponds to the magnetic field. 

The vector divergence is the $L_2$-adjoint of the gradient: $\operatorname{div} = - \operatorname{grad}^*$. Similarly, the covariant divergence is defined as $\operatorname{div}^{\nabla} = - \nabla^*$. The Bochner Laplacian on vector field is given by $\Delta \coloneqq  - \nabla^* \nabla = \operatorname{div}^{\nabla} \nabla$, and the nematic Bochner Laplacian is $\Delta_2 \coloneqq  - \nabla_2^* \nabla_2$. In this framework, the active shell model defined by Eqs.~\eqref{eq:displacementshell} without inertia and vanishing normal displacement ($w = 0$) takes the form:
\begin{subequations}\label{eq:displacementequations}
\begin{gather}
(\Delta + K + \tfrac{1+\nu}{1-\nu}\,\operatorname{grad} \circ \operatorname{div}) \bm {u} \notag \\
+ \tfrac{2\alpha{(1+\nu)}}{Yh} \operatorname{div}^\nabla( \mathcal{V}\Psi) =0\;,\\
\partial_{t} \Psi  = L \Delta_2 \Psi - \frac{1}{\epsilon^2}\left(|\Psi|^2-1\right) \Psi -\lambda \mathcal V^{-1} \traceless{\bm E}\;, 
\end{gather}
\end{subequations}
where $\epsilon$ is the coherence length controlling defect core size, and $\traceless{\bm E}$ is the traceless part of the symmetric strain tensor $\bm E =\delta \bm g / 2 = {1 \over 2}( \bm g \nabla \bm u + (\nabla \bm u)^\top \bm g )  \in \Gamma(\odot^2 T^*M)$, mapped to complex form via $\mathcal V^{-1}$. 
Strain tensor defines the Killing operator 
\begin{align}
    \mathcal K:  \Gamma(TM) \rightarrow \Gamma(\odot^2 T^*M), \quad \mathcal K: \bm u \mapsto \bm E.
\end{align} 
In our simulations the initialization of $\Psi$, corresponding to a disordered defect populated nematic sphere, are taken to be a normalized eigenmode of the nematic Bochner Laplacian $\Delta_2$, obtained by solving the eigenvalue problem $\Delta_2 \Psi_i = \lambda_i \Psi_i$. By the Rayleigh quotient, the lowest mode $\Psi_1$ minimizes the energy $\int_{\mathbb{S}^{2}}~\mathrm dA |\nabla \Psi|^2 $ under the $L_2$ constraint $\int_{\mathbb{S}^{2}}~\mathrm dA |\Psi|^2 = 1$. 
In our simulation, we use $i = 60$ and initialize the field as $\Psi|_{t = 0} = \Psi_{60} / |\Psi_{60}|$. 

\subsection{Discretization of surface and operators}

The surface $M$ is discretized as a triangular mesh $(\mathscr P, \mathscr E, \mathscr F)$, where $\mathscr P, \mathscr{F}, \mathscr{E}$ denote the sets of vertices, edges and faces. Local vector bases $\bm 1_f \in T_f M $ and $\bm 1_p \in T_p M$, along with their nematic counterparts $[\bm 1]_f \in T_f M / \mathbb Z_2 $ and $[\bm 1]_p \in T_p M / \mathbb Z_2$, are assigned to each face $f \in \mathscr F$ and vertex $p \in \mathscr{P}$. The displacement field $\bm u_f = \hat u_f \bm 1_f $ is discretized on faces, and the nematic field $\Psi_p = \hat \Psi_p \bm 1_p $ is discretized on vertices via complex scalars. These bases induce an angle-valued discrete Levi-Civita connection $\Omega_e$ on each edge $e \in \mathscr{E}$, which encodes the change of basis between adjacent element $i$ and $j$ as $\bm 1_j = e^{\mathrm i \Omega_e } \bm 1_i $, $[\bm 1]_j = e^{\mathrm i 2 \Omega_e } [\bm 1]_i $.

The discrete Gaussian curvature at each face $f$ measures the holonomy angle around that face:
\begin{align}
\textstyle
A_f K_f \coloneqq -  \sum_{e \prec f} \Omega_e  \mod 2 \pi \in (- \pi, \pi ]\;,
\end{align}
where the branch $(- \pi, \pi]$ for $A_f K_f$ is selected due to small domain area $A_f$.
On closed surfaces, this definition satisfies the discrete Gauss–Bonnet theorem $\sum_f A_f K_f = 2 \pi \chi$, with $\chi = |\mathscr P| - | \mathscr E| + | \mathscr F|$ being the Euler characteristics. 

The gradient $\operatorname{grad}: \Gamma_{\mathscr{P}}(M; \mathbb R) \rightarrow \Gamma_{\mathscr F}(TM)$ and divergence $ \operatorname{div}: \Gamma_{\mathscr F}(TM) \rightarrow \Gamma_{\mathscr P}(M; \mathbb R)$ follows standard linear conformaing finite element approximation. 
The covariant divergence $\operatorname{div}^{\nabla}: \Gamma_{\mathscr P}(TM \otimes TM) \rightarrow \Gamma_{\mathscr F}(T M)$ is discretized using a finite volume scheme with dihedral rotation from $T_p M$ to $T_f M$ as parallel transport. The vector Laplacian $\Delta = \star^{-1} d^\nabla \star \nabla: \Gamma_{\mathscr F}(TM) \rightarrow \Gamma_{\mathscr F}(TM)$ and nematic Laplacian $\Delta_2 =  \star^{-1} d^{\nabla_2} \star \nabla_2: \Gamma_{\mathscr P}(TM / \mathbb Z_2 ) \rightarrow \Gamma_{\mathscr P}(TM / \mathbb Z_2)$ are constructed using bundle-valued discrete exterior calculus. 

To incorporate the strain alignment $\lambda \mathcal{V}^{-1} \traceless{\bm E}$, we first interpolate the face-based displacement field $\bm u_f$ to vertices to obtain $\bm u_v$. A finite element gradient $\nabla: \Gamma_{\mathscr P}(M; \mathbb R^3) \rightarrow \Gamma_{\mathscr F}(T^*M \otimes \mathbb R^3)$ is then applied, precomposed with  dihedral rotation for parallel transport from $T_p M$ to $T_f M$. The face-based strain $\bm E_f = {1\over 2} ((\nabla \bm u)_f + (\nabla \bm u)^\top_f) $ is then interpolated to vertices as $\bm E_v$, made traceless, and mapped into Eq.~(\ref{eq:displacementequations}b) via the Veronese map. For further implementation details of these operators, we refer the reader to Ref.~\cite{Zhu:2025}. For temporal discretization, Eq.~(\ref{eq:displacementequations}b) is integrated using an implicit-explicit scheme: the alignment term $\lambda \mathcal{V}^{-1} \traceless{\bm E}$ is treated with explicit Euler, while the relaxation term $L \Delta_2 \Psi $ is handled via implicit Euler. Eq.~(\ref{eq:displacementequations}a) involves a linear Poisson solve.

\subsection{Variational formulation and Hodge-like decomposition}

On a closed, simply-connected manifold $M$, the space of smooth symmetric $(0, 2)$-tensor fields $\Gamma(\odot^2 T^*M)$ admits an $L_2$-orthogonal decomposition: 
\begin{align}
    \Gamma(\odot^2 T^*M) =  \mathcal E \oplus \mathcal{E}^\perp,
\end{align}
where 
\begin{align}
 \mathcal{E} := \{ \bm E = K \bm u \, \big| \, \bm u \in \Gamma(TM) \}
\end{align}
is the space of compatible strain fields (the image of the Killing operator), and 
\begin{align}
    \mathcal{E}^\perp = \{ \bm \Sigma \in \Gamma(\odot^2 T^*M) \, \big| \, \operatorname{div}^\nabla \bm \Sigma = 0 \}
\end{align}
is the space of self-equilibrated stress fields (kernel of the covariant divergence).
Orthogonality follows from integration by parts $\int \langle K \bm u, \bm \Sigma \rangle \,\mathrm{d} A  = -  \int \langle \bm u, \mathrm{div}^\nabla \bm \Sigma  \rangle  \,\mathrm{d} A= 0$.

The  strain field obtained from Eq.~(\ref{eq:displacementequations}a) is effectively the $L_2$-orthogonal projection of the active stress $\bm Q$ onto the compatible strain space $\mathcal E$,
\begin{align} 
\traceless{\bm E}  = -\frac{\alpha (1 + \nu)}{Yh} \mathbb P_{\mathcal E} \bm Q\;. 
\end{align}
Under the Veronese map $\mathcal{V}$, the corresponding projection in complex form is $\mathbb P_{\mathcal E}^\Psi = \mathcal V^{-1} \circ \traceless{~} \circ \mathbb P_{\mathcal E} \circ {\mathcal V}$, and its complement is $\mathbb P_{\mathcal E^\perp}^\Psi = 1 - \mathbb P_{\mathcal E}^\Psi$.

Using this projection, the displacement field $\bm u$ can be eliminated from  Eq.~\eqref{eq:displacementequations}, which can then be written in variational form:
\begin{align}
    \partial_t \Psi = -\frac{ \delta \mathcal F_\Psi}{\delta \Psi} =  -\frac{ \delta}{\delta \Psi}\,\int {\rm d}A\,F_\Psi\;,
\end{align}
with the energy density $F_\Psi$ given by 
\begin{align}
    F_\Psi = 
    \frac{L}{2} | \nabla_2 \Psi |^2 - \frac{\lambda \alpha (1 + \nu) }{2 Y h } |\mathbb P^\Psi_{\mathcal E} \Psi|^2 + \frac{1}{4 \epsilon^2} ( |\Psi|^2 - 1)^2.
\end{align}
In the long-wavelength regime where the alignment term $|\mathbb P^\Psi_{\mathcal E} \Psi|^2$ term dominates over diffusion $| \nabla_2 \Psi |^2$, the constraint imposed by the Ginzburg-Landau term, $\| \mathbb P_{\mathcal E}^\Psi \Psi \|^2 + \| \mathbb P_{\mathcal E^\perp}^\Psi \Psi \|^2 = \|\Psi \|^2 = \mathrm{const}$, forces a tradeoff between the strain-generating component $\| \mathbb P_{\mathcal E}^\Psi \Psi \|^2$ and self-equilibrated component $\| \mathbb P_{\mathcal E^\perp}^\Psi \Psi \|^2$. 
The system minimizes its energy by favoring the divergence-free (self-equilibrated) component when $\alpha \lambda < 0$, and the strain-generating component when $\alpha \lambda > 0$.

\section{\label{sec:shell}Derivation of Eqs.~\eqref{eq:shell}}

Eq.~(\ref{eq:shell}a) can be obtained by inverting Eq.~(\ref{eq:hooke}a) and applying to both sides the so-called incompatibility operator $\epsilon^{ik}\epsilon^{jk}\nabla_{k}\nabla_{l}$, with $\epsilon_{ij}$ the Levi-Civita tensor. For a generic rank-2 tensor $T_{ij}$, this gives
\begin{equation}\label{eq:tensor_compatibility}
\epsilon^{ik}\epsilon^{jl}\nabla_{k}\nabla_{l}T_{ij} = \nabla^{2}T-\nabla^{i}\nabla^{j}T_{ij}\;.	
\end{equation}
from which, taking Eq.~(\ref{eq:eom_appendix}a) into account, yields
\begin{equation}\label{eq:strain_compatibility}
\epsilon^{ik}\epsilon^{jl}\nabla_{k}\nabla_{l}E_{ij} = \frac{1}{Yh}\left[\nabla^{2}N+(1+\nu)\varphi\right]\;.
\end{equation}
Furthermore, as shown in Ref.~\cite{Giomi:2025}, the incompatibility of the strain tensor is related to the local variation of the Gaussian curvature by
\begin{equation}\label{eq:delta_K}
\epsilon^{ik}\epsilon^{jl}\nabla_{k}\nabla_{l}E_{ij} = -\delta K - \frac{E}{R^{2}}\;,
\end{equation}
where the trace $E$ of the strain tensor can be computed via Eq.~(\ref{eq:hooke}a) in the form
\begin{equation}\label{eq:strain_trace}
E = \frac{1-\nu}{Yh}\,N = -\frac{2(1-\nu)}{Yh}\,P^{({\rm e})}\;.
\end{equation}
Finally, combining Eqs.~\eqref{eq:strain_compatibility}, \eqref{eq:delta_K} and \eqref{eq:strain_trace} readily yields Eq.~(\ref{eq:shell}a). 

Eq.~(\ref{eq:shell}b), on the other hand, can be straightforwardly obtained upon replacing Eqs.~\eqref{eq:p_delta_k} in the homogeneous equation associated to Eq.~(\ref{eq:eom}b).

\section{\label{sec:green_function}Solutions of Eqs.~(\ref{eq:shell})}

In this Appendix we provide further details about the solution of Eqs.~(\ref{eq:shell}) in the condensed state, that is when the active flux given in Eq.~\eqref{eq:delta_flux}. To this end, we first recall that, on a sphere of radius $R$, the Dirac delta function attains the form
\begin{equation}
\delta(\bm{r}-\bm{r}')
= \frac{\delta(\cos\theta-\cos\theta')\delta(\phi-\phi')}{R^{2}}\;.
\end{equation}
In the present case, the spherical vector $\bm{r}'$ marks the the position of North and South pole, thus $\cos\theta'=\pm 1$ independently on $\phi$. Integrating Eq.~(\ref{eq:shell}a) over $\phi$ then gives
\begin{multline}\label{eq:helmholtz_appendix}
\left[\frac{1}{\sin\theta}\,\partial_{\theta}\,(\sin\theta\,\partial_{\theta})+(1-\nu)\right]P^{({\rm e})}\\
= \alpha(1+\nu)\left[\delta(\cos\theta-1)+\delta(\cos\theta+1)-1\right]\;.
\end{multline}
Both side of this equation can now be expanded in Legendre polynomials: i.e. $P^{({\rm e})}(\theta)=\sum_{\ell=0}^{\infty}a_{\ell}P_{\ell}(\cos\theta)$, with $a_{\ell}$ constants, and
\begin{equation}
\delta(\cos\theta-\cos\theta')
= \sum_{\ell=0}^{\infty}\frac{2\ell+1}{2}\,P_{\ell}(\cos\theta)P_{\ell}(\cos\theta')\;.
\end{equation}
Since $P_{0}(\cos\theta)=1$, the constant term on the right-hand side of Eq.~\eqref{eq:helmholtz_appendix} cancels the zero mode, so that $a_{0}=0$, whereas, for $\ell \ge 1$, one readily finds
\[
\frac{a_{\ell}}{\alpha(1+\nu)/2} = \left(\frac{2\ell+1}{2}\right)\frac{1+(-1)^{\ell}}{(1-\nu)-\ell(\ell+1)}\;, 	
\]
where we have used that $P_{\ell}(\pm 1)=(\pm 1)^{\ell}$. Because of this, $a_{\ell}=0$ for odd $\ell$ values, from which one recovers Eq.~\eqref{eq:symmetric_solution}. Similarly, Eq.~\eqref{eq:non_symmetric_solution} can be obtained by repeating the same derivation with the non-symmetric flux density given in Eq.~\eqref{eq:tau_flux}. For $\tau=1/2$ this reduces to Eq.~\eqref{eq:delta_flux}, while, for arbitrary $\tau$ values, repeating the previous steps yields
\begin{equation}\label{eq:a_tau}
\frac{a_{\ell}}{\alpha(1+\nu)/2} = \frac{\left(2\ell+1\right)[\tau+(-1)^{\ell}(1-\tau)]}{(1-\nu)-\ell(\ell+1)}\;. 	
\end{equation}
Introducing the parameter $\alpha_{\ell}$ defined in Eq.~\eqref{eq:alpha_l} finally gives Eq.~\eqref{eq:non_symmetric_solution}. 

If both stretching and bending are considered, parametrizing $\delta K = \sum_{\ell=0}^{\infty}b_{\ell}P_{\ell}(\cos\theta)$ in Eqs.~\eqref{eq:shell} gives the following expressions for the amplitudes $a_{0}$ and $b_{0}$:
\begin{subequations}
\begin{gather}
a_{0} =-\left(\frac{\gamma}{\gamma+1-\nu^{2}}\right)\frac{2P^{({\rm a})}+pR}{2}\;,\\
b_{0} =-\left(\frac{1-\nu}{\gamma+1-\nu^{2}}\right)\frac{2P^{({\rm a})}+pR}{D}\;.
\end{gather}
\end{subequations}
In the absence of active deviatoric stresses (i.e. $\alpha=0$, but $P^{({\rm a})}\ne 0$), these are the only non-vanishing terms in the expansion and, compared to the Eqs.~\eqref{eq:passive_deformations}, shows that the active pressure affect the overall stress balance by renormalizing the magnitude of the Laplace pressure $p/R$. For $\ell\ge 1$, standard algebraic manipulations gives
\begin{subequations}\label{eq:p_delta_k_legendre}
\begin{gather}
\frac{a_{\ell}}{D/2} = [1+\nu-\ell(\ell+1)]\,b_{\ell}\;,\\
\frac{b_{\ell}}{(1+\nu)/D}=-\frac{2\ell+1}{\gamma-\nu^{2}+[1-\ell(\ell+1)]^{2}}\,\alpha_{\ell}\;.
\end{gather}
\end{subequations}
where $\alpha_{\ell}$ is again defined in Eq.~\eqref{eq:alpha_l}. Notice that, in the limit $\gamma\to 0$, stretching and bending deformations decouple and the expression for $\alpha_{\ell}$ given in Eq.~(\ref{eq:p_delta_k_legendre}a) reduces to that of Eq.~\eqref{eq:a_tau}. 

\section{\label{sec:displacement}Derivation of Eqs.~\eqref{eq:apolar_w} and \eqref{eq:polar_w}}

In the apolar case, the normal displacement $w$ can be calculated directly from Eq.~(\ref{eq:p_delta_k}b), with $\delta K$ as a source term. Assuming again axial symmetry and taking $w=\sum_{\ell=0}^{\infty}c_{\ell}P_{\ell}(\cos\theta)$, readily gives
\begin{equation}\label{eq:c_l}
c_{\ell} = \frac{b_{\ell}}{\ell(\ell+1)-2}\;,\qquad \ell\ge 2\;,	
\end{equation}	
where $b_{\ell}$ is given in Eq.~(\ref{eq:p_delta_k_legendre}b). This expression, however, is ill-defined for $\ell=1$, as a consequence of $P_{1}(\cos\theta)=\cos\theta$ being in the kernel of the operator $\nabla^{2}+2/R^{2}$: i.e. 
\begin{equation}\label{eq:kernel}
\left(\nabla^{2}+\frac{2}{R^{2}}\right)w_{1} = 0\;,
\end{equation}
where $w_{1}=c_{1}\cos\theta$. To calculate $c_{1}$ we used Eqs.~(\ref{eq:displacementshell}a) and (\ref{eq:displacementshell}b) directly and, taking advantage of Eq.~\eqref{eq:kernel}, write 
\begin{subequations}\label{eq:u1_w1}
\begin{gather}
\frac{1+\nu}{1-\nu}\,\nabla\left(\nabla\cdot \bm{u}_{1}+\frac{2w_{1}}{R}\right) \notag \\
+\left(\nabla^2 + \frac{1}{R^2} \right)\bm{u}_{1}+\frac{2(1+\nu)}{Yh}\bm{f}_{1}=\bm{0}\;,\\
\nabla\cdot \bm{u}_{1}+\frac{2w_{1}}{R} = 0\;,
\end{gather}
\end{subequations}
where $\bm{u}_{1}=u_{10}^{(1)}\bm{\Psi}_{1}^{0}+u_{10}^{(2)}\bm{\Phi}_{1}^{0}$. Next, using the properties vector spherical harmonics summarized in Appendix~\ref{sec:active_force}, it is possible to derive from Eqs.~\eqref{eq:u1_w1} the following relations for the coefficients $u_{10}^{(1)}$, $u_{10}^{(2)}$ and $w_{1}$. That is
\begin{subequations}
\begin{gather}
u_{10}^{(1)} = w_{1} = \frac{2R^{2}(1+\nu)}{Yh}\,f_{1}^{(1)}\;,\\
u_{10}^{(2)} = \frac{2R^{2}(1+\nu)}{Yh}\,f_{1}^{(2)}\;,
\end{gather}
\end{subequations}	
form which one readily obtain Eq.~\eqref{eq:polar_w}. It must be noted that $\bm{\Psi}_{1}^{0}=-\sqrt{3/(4\pi)}\,\sin\theta\,\bm{e}_{\theta}$, hence $f_{1}^{(1)} \sim -\alpha_{1}=-\alpha(2\tau-1)$.
\bibliography{hydra_arxiv.bbl}
\end{document}